\documentclass[twocolappendix, numberedappendix]{emulateapj}
\usepackage{textcomp}
\usepackage{amsmath}
\usepackage{color}

\def\MgII{Mg\,{\sc ii}}
\def\O3{[O\,{\sc iii}]}
\def\Hbeta{$\rm{H}\beta$}

\shorttitle{SDSS-RM: Variability of Quasar BELs}

\begin{document}
\slugcomment{\bf Draft: \today}
\title{The Sloan Digital Sky Survey Reverberation Mapping 
Project: Ensemble Spectroscopic Variability of Quasar 
Broad Emission Lines}

\author{ Mouyuan Sun\altaffilmark{1,2} }
\author{ Jonathan R. Trump\altaffilmark{1,$\dagger$} }
\author{ Yue Shen\altaffilmark{4,5,$\dagger$} }
\author{ W. N. Brandt\altaffilmark{1,3} }

\author{ Kyle Dawson\altaffilmark{6} }
\author{ Kelly D. Denney\altaffilmark{7,$\ddagger$} }
\author{ Patrick B. Hall\altaffilmark{8} }
\author{ Luis C. Ho\altaffilmark{4,9} }
\author{ Keith Horne\altaffilmark{10} }
\author{ Linhua Jiang\altaffilmark{4} }
\author{ Gordon T. Richards\altaffilmark{11} }
\author{ Donald P. Schneider\altaffilmark{1} }

\author{ Dmitry Bizyaev\altaffilmark{12,13} }
\author{ Karen Kinemuchi\altaffilmark{12} }
\author{ Daniel Oravetz\altaffilmark{12} }
\author{ Kaike Pan\altaffilmark{12} }
\author{ Audrey Simmons\altaffilmark{12} }

\altaffiltext{1}{
  Department of Astronomy \& Astrophysics and Institute for
  Gravitation and the Cosmos, 525 Davey Lab, The Pennsylvania State
  University, University Park, PA 16802, USA}
\altaffiltext{2}{
  Department of Astronomy and Institute of Theoretical 
  Physics and Astrophysics, Xiamen University, Xiamen, Fujian 361005, China}  
\altaffiltext{3}{
  Department of Physics, The Pennsylvania State University, University Park, PA 16802, USA}
\altaffiltext{4}{
  Kavli Institute for Astronomy and Astrophysics, Peking University, Beijing 100871, China}
\altaffiltext{5}{
  Carnegie Observatories, 813 Santa Barbara Street, Pasadena, CA 91101, USA}
\altaffiltext{6}{
  Department of Physics and Astronomy, University of Utah, Salt Lake City, UT 84112}
\altaffiltext{7}{
  Department of Astronomy, The Ohio State University, 140 West 18th Avenue, Columbus, 
  OH 43210, USA}
\altaffiltext{8}{
  Department of Physics and Astronomy, York University, Toronto, ON M3J 1P3, Canada}
\altaffiltext{9}{
  Department of Astronomy, School of Physics, Peking University, Beijing 100871, China}
\altaffiltext{10}{
  SUPA Physics/Astronomy, Univ. of St. Andrews, St. Andrews KY16 9SS, Scotland, UK}
\altaffiltext{11}{
  Department of Physics, Drexel University, 3141 Chestnut St., Philadelphia, PA 19104, USA}
\altaffiltext{12}{Apache Point Observatory and New Mexico State
University, P.O. Box 59, Sunspot, NM, 88349-0059, USA}
\altaffiltext{13}{Sternberg Astronomical Institute, Moscow State
University, Moscow}
\altaffiltext{$\dagger$}{Hubble Fellow}
\altaffiltext{$\ddagger$}{NSF Astronomy and Astrophysics Postdoctoral Fellow}

\begin{abstract}
We explore the variability of quasars in the \MgII \ and \Hbeta \ broad 
emission lines and UV/optical continuum emission using the Sloan Digital 
Sky Survey Reverberation Mapping project (SDSS-RM). This is the largest 
spectroscopic study of quasar variability to date: our study includes 
$29$ spectroscopic epochs from SDSS-RM over $6$ months, containing 
$357$ quasars with \MgII \ and $41$ quasars with \Hbeta . On longer 
timescales, the study is also supplemented with two-epoch data from 
SDSS-I/II. The SDSS-I/II data include an additional $2854$ quasars with 
\MgII \ and $572$ quasars with \Hbeta . The \MgII \ emission line is 
significantly variable ($\Delta f/f \sim 10\%$ on $\sim100$-day timescales), 
a necessary prerequisite for its use for 
reverberation mapping studies. The data also confirm that continuum 
variability increases with timescale and decreases with luminosity, and the 
continuum light curves are consistent with a damped random-walk model 
on rest-frame timescales of $\gtrsim 5$~days. We compare the emission-line 
and continuum variability to investigate the structure of the broad-line region. 
Broad-line variability shows a shallower increase with timescale compared to 
the continuum emission, demonstrating that the broad-line transfer function 
is not a $\delta$-function. \Hbeta \ is more variable than \MgII \ (roughly 
by a factor of $\sim 1.5$), suggesting different excitation mechanisms, optical 
depths and/or geometrical configuration for each emission line. The ensemble 
spectroscopic variability measurements enabled by the SDSS-RM project have 
important consequences for future studies of reverberation mapping and black 
hole mass estimation of $1<z<2$ quasars. 
\end{abstract}


\section{introduction}
Aperiodic luminosity variations across the electromagnetic spectrum are 
an ubiquitous feature of quasars\footnote{We use the term ``quasar'' to 
generically refer to active galactic nuclei with optical broad emission lines, 
regardless of luminosity.}  \citep[for a review, see][]{ulrich97}. The optical 
and ultraviolet (UV) continuum emission of a typical (nonblazar) quasar 
vary by a few tenths of a magnitude on timescales from weeks to years. 
Theoretically, the observed quasar continuum variability may be driven 
by several kinds of complex instabilities in the accretion disk 
\citep[e.g.,][]{lyu97, czerny99, czerny06, li08, kelly11}. Observationally, 
however, photometric light curves can be well-modeled by a simple stochastic 
process: the damped random-walk (DRW) model (for the statistical properties 
of the DRW model, see Section~\ref{sec:drw}). The large body of work on 
both individual and ensemble quasar variability has established that the 
amplitude of continuum variability increases with time between epochs, 
decreases with quasar luminosity and rest-frame wavelength, and is 
independent of redshift \citep[e.g.,][]{uom76, hook94, giv99, haw02, vb04, 
dv05, bau09, kelly09, mac12, zuo12}. 

Quasar broad emission lines arise from \mbox{doppler-broadened} line emission 
from gas deep within the gravitational potential well of the supermassive black 
hole, i.e., the broad line region (BLR), that is photoionized by the extreme UV 
(EUV) accretion disk continuum radiation.  As a result, they vary in response to 
the continuum variations after a light-travel time delay. The amplitude and shape 
of the emission-line response are governed by the broad-line transfer 
function \citep{bla82}. The transfer function is ultimately determined 
by the radiative mechanism, as well as the structure, dynamics, and 
ionization state of the BLR. The variability time delay enables reverberation 
mapping to study the structure of the BLR \citep[e.g.,][]{gas09} and (with 
assumptions about the geometry and BLR dynamics) estimate the black hole mass 
\citep[e.g.,][]{pet93, lao98, pet98, kas00, pet14}. In theory, the reverberation mapping 
technique can be performed using any broad emission lines that respond to the 
variability of continuum emission. In practice, however, reverberation mapping has 
been largely restricted to using the \Hbeta \ emission line in low-luminosity systems 
at low redshift \citep[e.g.,][]{pet06}.

Employing reverberation mapping with optical spectroscopy at $z>1$ is critical 
for our understanding of the mass growth of supermassive black holes, as most 
mass growth occurs at this epoch. However, optical reverberation mapping at 
$z>1$ requires using rest-frame UV broad emission lines, such as \MgII . \MgII \ 
is ionized by $E>15\ \rm{eV}$ photons (i.e., the ionization energy to go from 
\MgII \ to Mg\,{\sc iii}), which is similar to the ionization energy ($E>13.6\ 
\rm{eV}$) of H\,{\sc i} (which subsequently leads to recombination and 
\Hbeta \ emission). In addition, the similarity between the \MgII \ and \Hbeta \ 
velocity widths indicates the two lines are produced in a similar environment 
and distance from the central black hole \citep[e.g.,][]{shen08, wan09, shen12}. 
However, the validity of \MgII \ as a black-hole mass estimator is under debate. 
The variability of \MgII \ has only been observed in a handful of quasars, and 
it is not entirely clear that this variability can be robustly traced to coherent 
reverberation of the ionizing EUV continuum \citep[e.g.,][]{cla91,rei94, tre07, 
woo08, hry14, cac15}. Moreover, the radiative mechanism to produce \MgII \ 
may also differ from that for \Hbeta, as the former may mostly be collisionally 
excited while the latter is a recombination line \citep[e.g.,][]{mac72, net80}. 
Therefore, it is vital to investigate the variability of \MgII \ for a large 
sample. This can only be accomplished with a large multi-epoch broad line 
quasar spectroscopic survey. 

In this work, we measure the ensemble variability of broad emission lines 
and continuum emission using data from the Sloan Digital Sky Survey 
Reverberation Mapping project \citep[SDSS-RM,][]{shen15} and from 
supplemental observations in SDSS-I and SDSS-II. In Section~\ref{sec:sample} 
we describe the SDSS observations. Section~\ref{sec:lcv} presents 
the basic variability of quasar light curves. In Section~\ref{sec:obsvar}, 
we present the distribution of the observed luminosity variability. In 
Section~\ref{sec:sfall}, we introduce the structure function as a tool 
to study quasar variability on different timescales. Section~\ref{sec:sfcont} 
presents the structure function of continuum emission, and 
Section~\ref{sec:sfeml} describes the structure function of \MgII \ and 
\Hbeta . In Section~\ref{sec:disall}, we discuss the physical implications 
of our results. The main results of this work are summarized 
in Section~\ref{sec:summary}. We adopt a flat $\Lambda$CDM cosmology with 
$\Omega_{\rm{m}}=0.3$ and $h_0=0.7$. Throughout this work, ``$\left<x\right>$'' 
and ``$\tilde{x}$'' represent the arithmetic mean and the median of the 
variable $x$, respectively.

\section{SAMPLE DEFINITION}
\label{sec:sample}
In this work, we use SDSS data to study quasar variability. The SDSS 
$2.5$~m telescope is described by \cite{gun06}. \cite{eis11} give a 
technical summary of the SDSS-III project, and the SDSS/BOSS 
spectrograph and reduction pipeline are described by 
\cite{bol12, daw13, sme13}. 

We focus on the data from the SDSS-RM project, which is an ancillary 
program within SDSS-III and probes the variability of quasars on 
rest-frame timescales of $1\lesssim\Delta t \lesssim 100\ \rm{days}$. 
The SDSS-I/II projects provide ancillary data to study the variability 
of quasars on rest-frame timescales of $100\lesssim \Delta t~\lesssim 
1000\ \rm{days}$. 

\subsection{SDSS-RM quasars}
\label{sec:rm-sample}
The SDSS-RM sample, observed during the \mbox{SDSS-III} BOSS survey 
\citep{eis11, daw13}, consists of $849$ quasars, each with $32$ epochs 
of observations (with a total exposure time of $\sim 60$ hours): for 
technical details, see \cite{shen15}. Three out of the $32$ epochs 
have low S/N spectra (i.e., $\rm{S/N}<0.7\left< \rm{S/N}\right>$) and 
are rejected. The spectrograph has a wavelength range 
of $3650$--$10400\ \rm{\AA}$ with a spectral resolution of $R \sim 2000$ 
\citep{sme13}. The flux calibration was performed based on $70$ standard 
stars at each epoch, using an improved version of the standard BOSS 
pipeline \citep[for more details, see][]{shen15}. 

For each quasar, we fit the quasar spectra and obtained (depending on 
the observed spectral coverage) $L=\lambda L_{\lambda}$ for 
rest-frame\footnote{Throughout this work, the wavelengths 
of quasar features are always rest-frame, unless otherwise specified.} 
$\lambda=3000\ \rm{\AA}$ and/or $\lambda= 5100\ \rm{\AA}$, as well 
as the luminosities and the full-width-half-max velocity ($v_{\rm{FWHM}}$) 
of \MgII \ and/or \Hbeta . The details of the continuum and the line fitting 
are described in detail in earlier work \citep[see][]{shen08, shen11}. 

We then selected a parent sample of $731$ quasars with broad \MgII \ 
or \Hbeta \ by requiring $z\le 2.462$. This requirement ensured that either 
\MgII \ or \Hbeta \ (or both) and their respective continuum regions 
were present in the BOSS spectrum. We further required that the median 
$v_{\rm{FWHM}}$ over $29$ epochs $\widetilde{v}_{\rm{FWHM}}>1000\ 
\rm{km\ s^{-1}}$. 

The observed variability is a superposition of the intrinsic variability 
of quasars, the measurement errors, and the spectrophotometric errors 
(see Appendix~\ref{sec:sperr}). We applied the following sample-selection 
criteria to obtain an unbiased measurement of the intrinsic variability 
of quasars: 
\begin{itemize}
  \item There are ``dropped'' epochs where the fiber was not properly 
  plugged into the spectroscopic mask resulting in sudden, unusually 
  large reduction in the spectral flux \citep{shen15}. To avoid these 
  ``dropped'' spectra, we reject epochs with $\left|m-\tilde{m}\right|>1$, 
  where $m$ is the magnitude at a given epoch and $\tilde{m}$ is the 
  median magnitude over $29$ epochs. About $1\%$ of the total epochs 
  are excluded \citep[roughly consistent with][]{shen15}. Visual 
  inspection shows that this criterion does not reject any cases of 
  real variability.
  
  \item There are spectra with low $\rm S/N$ in the $r$-band and 
  therefore with poor flux calibration, since the spectra are 
  calibrated only in $r$-band. To identify these quasars, 
  we reject quasars with $r$-band $\rm S/N$ less than $50$, 
  measured by convolving the spectra with the SDSS $r$-band filter. 
  
  \item There are spectra that have strong broad-absorption features 
  in their \MgII \ profiles. In this case, the measurements of 
  \MgII \ flux can be problematic. To avoid this issue, we reject 
  these quasars. 
  
  \item The observed-frame quasar spectra are dominated 
  by sky lines at wavelengths larger than $9000\ \rm{\AA}$. To avoid 
  this issue, we reject quasars with redshift $z>2.0$ for the $3000\ 
  \rm{\AA}$ continuum and \MgII \ and $z>0.8$ for the $5100\ \rm{\AA}$ 
  continuum and \Hbeta . 

  \item There are quasars whose observed variability is dominated by 
  measurement errors rather than intrinsic variability. In this case, 
  the estimated intrinsic variability can be highly biased. To avoid 
  this issue, we reject quasars with median signal-to-noise ratio 
  ($\widetilde{\rm S/N}$) of the continuum or line luminosity $<10$, 
  i.e., $\widetilde{\rm S/N}<10$. 
  
  \item There are two sources with nearby (angular distance $<5^{''}$, 
  using the SDSS imaging) bright foreground object that contaminate 
  the quasar spectra in some epochs and introduce artificial flux 
  variations in the fibers. To avoid this issue, we reject these two 
  sources. 
\end{itemize}
The final sample (i.e., clean sample) that passed the selection criteria 
and will be used for subsequent variability analysis of each continuum 
and emission-line component is summarized as follows:
\begin{itemize}
  \item the $3000\ \rm{\AA}$ continuum: $577$ quasars;
  
  \item \MgII\ broad emission line: $357$ quasars;
  
  \item the $5100\ \rm{\AA}$ continuum: $97$ quasars; 
  
  \item \Hbeta \ broad emission line: $41$ quasars. 
\end{itemize}

For each quasar in the parent sample, we estimated $L_{\rm Bol}$ 
using $L_{3000}$ and/or $L_{5100}$. The bolometric correction 
factor is assumed to be $5$ for $L_{3000}$ and $10$ for $L_{5100}$ 
\citep[e.g.,][]{ric06}\footnote{Monochromatic bolometric correction 
factors are likely to be luminosity-dependent \citep[e.g.,][]{lus12, kra13}. 
In this work, we use $L_{\rm Bol}$ only to divide the sources into 
different luminosity bins, and so we adopt the constant bolometric 
corrections of \cite{ric06} for simplicity. Our conclusions do not 
change if we instead adopt luminosity-dependent bolometric 
correction factors, which merely change the bin divisions by a 
small amount.}. If a spectrum covered both $3000\ \rm{\AA}$ 
and $5100\ \rm{\AA}$ we adopted the $L_{3000}$ estimator 
which is less contaminated by host galaxy starlight and 
therefore provides a less biased measure of quasar 
luminosity. 

We also measured $M_{\rm{BH}}$ using the single-epoch broad-line 
(\MgII \ and/or \Hbeta ) estimators: 
\begin{equation}
\label{eq:mse}
\log (\frac{M_{\rm{BH}}}{M_{\odot}}) = A + B\log (\frac{\lambda 
L_{\lambda}}{10^{44}\ \rm{erg\ s^{-1}}}) + 2\log 
(\frac{v_{\rm{FWHM}}}{1000\ \rm{km\ s^{-1}}}) 
\end{equation}
For \Hbeta , $\lambda = 5100\ \rm{\AA}$, $A=6.91$, $B=0.5$ 
\citep{ves06}; for \MgII, $\lambda = 3000\ \rm{\AA}$, $A=6.74$, 
$B=0.62$ \citep{shen11}. The quantities $\lambda L_{\lambda}$ and 
$v_{\rm{FWHM}}$ are measured from the averages of $29$ epochs. 
The measurement uncertainties of $\lambda L_{\lambda}$ and 
$v_{\rm{FWHM}}$ are typically $0.01$ dex and $5\%$, respectively. 
The uncertainty of $M_{\rm BH}$ is therefore dominated by the 
intrinsic uncertainty of the estimator, which is $\sim 0.4$ dex 
\citep[for a recent review, see][]{shen13}. If a spectrum included 
both \MgII \ and \Hbeta , we adopted the average $M_{\rm BH}$ 
(there are $134$ such spectra in the parent samples), following 
\cite{ves09}. For these 
sources, the difference of the two $M_{\rm BH}$ estimators has a 
median value of $-0.12$ dex (i.e., on average, \MgII \ estimators 
give slightly higher $M_{\rm BH}$ than those using \Hbeta ) and a 
standard deviation of $0.35$ dex. Only for three sources in the 
parent samples (but not in the clean samples) are the 
differences in the two $M_{\rm BH}$ estimators larger than $1$ dex. 
Two are $z<0.4$ sources and therefore $v_{\rm FWHM}$ of \MgII \ 
is not well constrained; the redshift of the remaining source is 
$z=0.923$ and therefore $v_{\rm FWHM}$ of \Hbeta \ is poorly 
measured. 

\begin{figure}
\epsscale{1.2}
\plotone{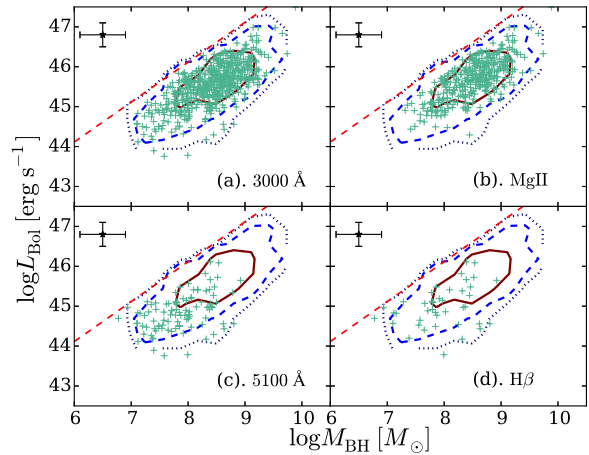}
\caption{The distribution of the parent sample of SDSS-RM sources 
in the $L_{\rm{Bol}}-M_{\rm{BH}}$ plane. The solid, dashed and 
dotted contours in each panel show boundaries of regions 
containing $68.2\%$, $95.4\%$ and $99.7\%$ of the total 
number of sources in our parent SDSS-RM sample. Points 
in each panel represent the SDSS-RM quasars that passed 
the selection criteria (i.e., the clean sample). The red 
dashed line in each panel represents the Eddington luminosity 
as a function of $M_{\rm BH}$. The error bar in each panel indicates 
the typical uncertainty of quasar luminosity (due to the uncertainty 
of the bolometric correction) and $M_{\rm BH}$.
}
\label{fig:rm-sample}
\end{figure}

Figure~\ref{fig:rm-sample} presents the distributions of the parent 
sample and the clean sample in the $L_{\rm{Bol}}-M_{\rm{BH}}$ plane 
for each component. For the $3000\ \rm{\AA}$ continuum and \MgII , 
the clean samples and the parent samples share similar parameter space. 
However, the clean $5100\ \rm{\AA}$ continuum and \Hbeta \ samples 
cover only the low-luminosity and small $M_{\rm BH}$ portion of parameter 
space, compared to the parent samples. This result is due to the fact that we 
can only measure the $5100\ \rm{\AA}$ continuum and \Hbeta \ for $z<0.8$ 
sources and low-redshift sources are more likely to be less luminous, 
on average, for a flux-limited survey. 

\subsection{SDSS-I/II quasars}
We use ancillary data compiled from the SDSS-I/II surveys 
\citep{sdss} to measure the variability of quasars on longer 
timescales. This sample includes every quasar with multiple 
spectroscopic epochs (at least two) in SDSS-I/II. Most of 
these SDSS-I/II quasars were only observed twice. Only a 
small fraction of these quasars are observed with SDSS-III. 
We do not supplement this set of quasars with SDSS-III 
observations due to the difference in flux calibration from 
SDSS-I/II to SDSS-III \citep{mar15}. We fit the spectra 
and obtained the luminosities at $\lambda=3000\ \rm{\AA}$ 
and $\lambda=5100\ \rm{\AA}$ and also the luminosities of 
the emission lines, \MgII \ and \Hbeta , following our 
earlier work \citep{shen08, shen11}. For the $3000\ \rm{\AA}$ 
continuum and \MgII , a parent sample consisting of $4599$ 
quasars was compiled. For the $5100\ \rm{\AA}$ continuum 
and \Hbeta , a parent sample with $1347$ quasars was 
constructed. 

\begin{figure}
\epsscale{1.2}
\plotone{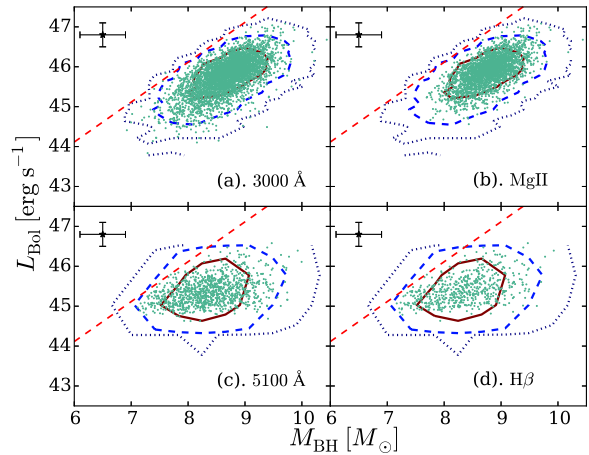}
\caption{The distribution of our SDSS-I/II sources in the 
$L_{\rm{Bol}}-M_{\rm{BH}}$ plane. The solid, dashed and 
dotted contours in each panel show boundaries of regions 
containing $68.2\%$, $95.4\%$ and $99.7\%$ of the total 
number of sources in our initial SDSS-I/II sample. Points 
in each panel represent the distribution of our final 
SDSS-I/II sample after applying our selection criteria. 
The red dashed line in each panel represents the 
Eddington luminosity as a function of $M_{\rm BH}$. The 
error bar in each panel indicates the typical uncertainty of 
quasar luminosity and $M_{\rm BH}$.
}
\label{fig:12-sample}
\end{figure}

To ensure that the intrinsic variability is accurately 
measured, we applied the fourth and fifth selection 
criteria described in Section~\ref{sec:rm-sample} 
to our SDSS-I/II sources. It is not necessary to remove 
dropped spectra or constrain the $r$-band S/N, since 
these problems are less frequent in SDSS-I/II spectra 
(since the spectroscopic flux limit is much shallower). 
Even if these problems occur in a epoch for a quasar, 
it only affects a single flux pair (i.e., $f_2/f_1$) in 
contrast to SDSS-RM, in which it affects $28$ flux pairs, 
and we use hundreds to thousands of flux pairs 
when calculating the variability. The final sample (i.e., 
clean sample) that passed the selection criteria and will 
be used for subsequent variability analysis of each 
continuum and emission-line component is summarized 
as follows:
\begin{itemize}
  \item the $3000\ \rm{\AA}$ continuum: $4213$ quasars;
  
  \item \MgII \ broad emission line: $2844$ quasars;
  
  \item the $5100\ \rm{\AA}$ continuum: $1064$ quasars;
  
  \item \Hbeta \ broad emission line: $572$ quasars.
\end{itemize}
For each quasar in the parent samples, we calculated $M_{\rm{BH}}$ 
and $L_{\rm Bol}$ following the methods described in 
Section~\ref{sec:rm-sample}. Figure~\ref{fig:12-sample} shows 
the distribution of the SDSSI/II sources in the $L_{\rm{Bol}}-M_{\rm{BH}}$ 
plane. Three contours again indicate the area that contains $68.27\%$, 
$95.45\%$ and $99.73\%$ of the parent samples. The points display 
the distributions of the final samples. 
For each component, the final sample covers similar parameter space 
as the parent SDSS-I/II sample. For the $3000\ \rm{\AA}$ continuum 
and \MgII , compared to the distribution of final SDSS-RM sources 
considered, most of the final SDSS-I/II sources  cover a similar 
parameter space. For the $5100\ \rm{\AA}$ continuum and \Hbeta , 
compared to the distribution of the final SDSS-RM sources, the final 
SDSS-I/II sources cover the high-luminosity and large $M_{\rm BH}$ 
parameter space. Our analysis treats the SDSS-RM and SDSS-I/II datasets 
independently, rather than combining the two datasets. This approach 
was adopted also because the two datasets have different time resolution 
and the variability of broad emission lines may be correlated with 
additional parameters beyond $L_{\rm{Bol}}$ and $M_{\rm{BH}}$. 

\section{Light Curve Study}
\label{sec:lcv}
\subsection{Observed Light Curves}
\begin{figure}
\epsscale{1.2}
\plotone{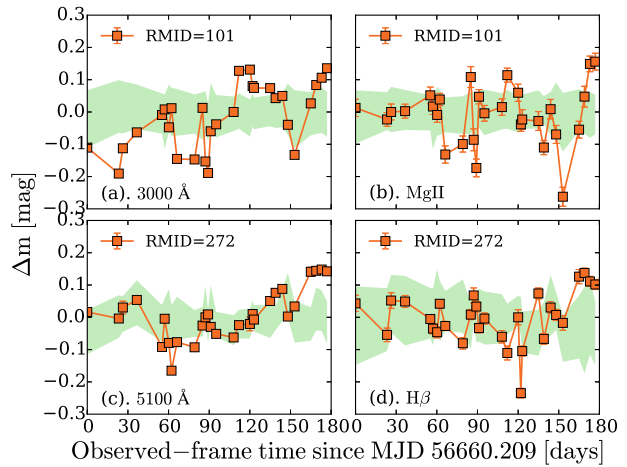}
\caption{Continuum and emission-line light curves of typical SDSS-RM 
sources (red lines/points), and the $25$--$75$ percentile range 
of the ensemble light curves (green shaded regions). The observed 
ensemble variability is a few tenths of a magnitude in both the continuum 
and broad emission lines. 
}
\label{fig:lcv}
\end{figure}

We start with the basic properties of the observed light 
curves. Figure~\ref{fig:lcv} plots continuum and broad 
emission-line light curves of the clean SDSS-RM samples. 
Throughout this work, we use magnitude (rather than flux or 
luminosity) changes to characterize variability so that it is 
straightforward to compare our results with previous photometric 
studies. The light curves do indicate that there is variability 
in both continuum emission and broad emission lines. However, 
the observed variability is a superposition of intrinsic 
variability, measurement errors, and spectrophotometric 
errors. The spectrophotometric errors of SDSS-RM for a single 
epoch are at least $\sim 0.04\ \rm{mag}$, and larger at long 
and short wavelengths, as computed from standard stars: see 
Appendix~\ref{sec:sfstar}. 

\subsection{Basic Light-Curve Variability}
\label{sec:lcvar}
Following \cite{ses07}, we define the basic intrinsic variability 
of a light curve with $n$ epochs as, 
\begin{equation}
\sigma_{lc}=\sqrt{\frac{1}{n-1} \sum_{i=1}^{n}(m_i-\left<m\right>)^2 
- \left<\sigma^2_{\rm{e}}\right>)}
\end{equation}
where $m_i$ is the observed magnitude at each epoch, its 
corresponding uncertainty is $\sigma_{\rm{e}}$ ($\sigma_{\rm{e}}$ 
is a summation of measurement errors and spectrophotometric 
errors in quadrature), and $\left<m\right>$ is the mean magnitude 
across all epochs. $\sigma_{lc}$ is set to zero if the operand 
within the square root is less than zero (the fraction of such sources 
is less than $10\%$). This is motivated by the fact 
that the variance of the 
quasar variability is $\geq 0$. The median ratios between the 
observed variance and the variance due to error (measurement and 
spectrophotometric), $\frac{1}{n-1} \sum_{i=1}^{n}(m_i-\left<m\right>)^2/ 
\left<\sigma^2_{\rm{e}}\right>$, is $5.6$ for the $3000\ \rm{\AA}$ 
continuum, $2.5$ for \MgII , and $2.9$ for \Hbeta . Note that we 
required $\widetilde{\rm S/N}\geq 10$ to ensure that, in most cases, 
the measurement errors do not dominate the observed variances.

\subsubsection{The $3000$ \rm{$\rm{\AA}$} \it{Continuum}}
\begin{figure}
\epsscale{1.15}
\plotone{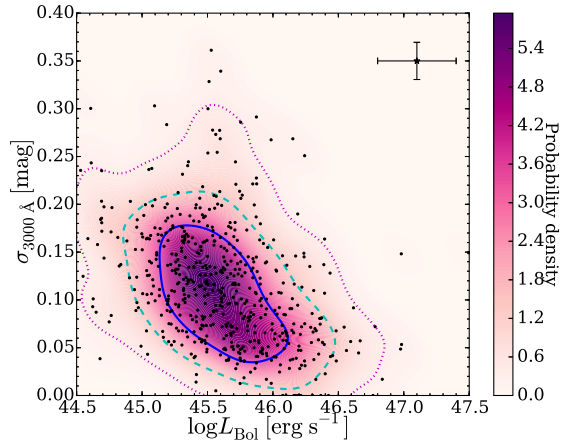}
\caption{The light-curve variability of the $3000\ \rm{\AA}$ continuum. 
Each point represents a quasar. The solid, dashed, and dotted lines 
represent $1\sigma$, $2\sigma$ and $3\sigma$ bounds of the quasar 
number distribution. The color indicates the probability density of each 
point (calculated via kernel density estimation). The light curve variability 
decreases with quasar luminosity. In this figure and 
Figures~\ref{fig:mgii-lcvar}-\ref{fig:hbeta-mgii-lcvar}, the error bars 
indicate the median uncertainty of the light-curve variability. }
\label{fig:l3k-lcvar}
\end{figure}

We first present the variability of the $3000\ \rm{\AA}$ continuum as 
a function of quasar luminosity (Figure~\ref{fig:l3k-lcvar}). We tested 
the correlation between the light-curve variability and $L_{\rm{Bol}}$ 
using the Spearman rank correlation test\footnote{We also applied 
the Kendall rank correlation to test our data. The results of this test 
are consistent with those of the Spearman rank test, unless otherwise 
specified.}. The null hypothesis of this 
test is that there is no correlation between the input datasets. The 
correlation coefficient of this test is $\rho=-0.44$. The $p$ value 
(i.e., the probability of being incorrect in rejecting the null hypothesis) 
is only $5\times 10^{-29}$. Therefore, we conclude that there is a 
significant anti-correlation between the light-curve variability 
and quasar luminosity at significance level of 
$\alpha = 0.01$\footnote{Throughout this work, we adopt $\alpha=0.01$ 
when we perform statistical hypothesis tests.} (i.e., the probability 
threshold below which the null hypothesis will be rejected). As we 
show in Section~\ref{sec:sfcont}, this anti-correlation is not simply 
due to more luminous quasars having smaller $\sigma_{\rm e}$ 
or typically being at higher redshift (sampling shorter rest-frame timescales). 
This behavior is consistent with previous work based on broad-band 
photometric data \citep[e.g.,][]{vb04, dv05, bau09, mac12}. 

Previous work suggests that the Eddington ratio, which 
depends on both quasar luminosity and $M_{\rm BH}$, is the main 
driver of quasar continuum variability \citep[e.g.,][]{wil08, bau09, 
ai10, mac10}. We verify this scenario by testing the possible correlation 
between the light-curve variability of the $3000\ \rm{\AA}$ continuum 
and the Eddington ratio. We stress that the uncertainty of the Eddington 
ratio is rather large since both quasar luminosity and $M_{\rm BH}$ are 
uncertain by a factor of $\sim 3$. The Spearman correlation 
coefficient of this test is $\rho=-0.2$ and the $p$ value is $5\times 10^{-7}$. 
This anti-correlation is consistent with previous work.

\subsubsection{\rm{\MgII}}
\begin{figure}
\epsscale{1.15}
\plotone{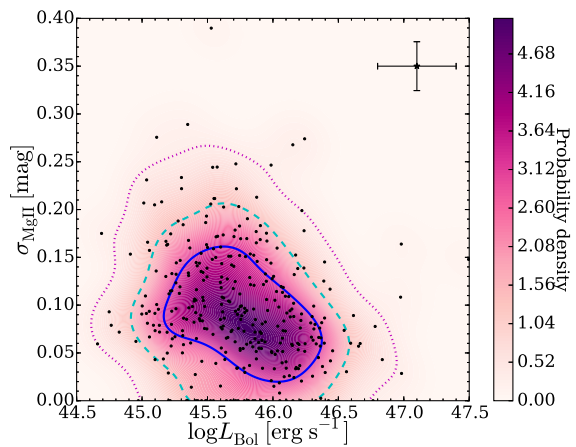}
\caption{The light-curve variability of \MgII . Each point 
represents a quasar. The solid, dashed, and dotted lines 
represent $1\sigma$, $2\sigma$ and $3\sigma$ bounds of the 
quasar number distribution. The color indicates the probability 
density of each point. The light curve variability of \MgII \ 
also decreases with quasar luminosity. }
\label{fig:mgii-lcvar}
\end{figure}

We explore the variability of the light curve of \MgII \ as a function 
of quasar luminosity (Figure~\ref{fig:mgii-lcvar}). 
With the same Spearman correlation test, for \MgII , $\rho=-0.24$ 
and the $p$ value is $3\times10^{-6}$. Therefore, again there is 
an anti-correlation between the light-curve variability of \MgII \ and 
quasar luminosity. The same Spearman correlation test also 
suggests that the light-curve variability of the \MgII \ and the Eddington 
ratio are anti-correlated (with $\rho=-0.24$, and the $p$ value is 
$7\times 10^{-6}$). As we show in Section~\ref{sec:sfcont}, these 
anti-correlations are not simply due to more luminous quasars preferentially 
having smaller $\sigma_{\rm e}$ or lying at higher redshift (with 
shorter rest-frame timescales).

\begin{figure*}
\epsscale{1.15}
\plotone{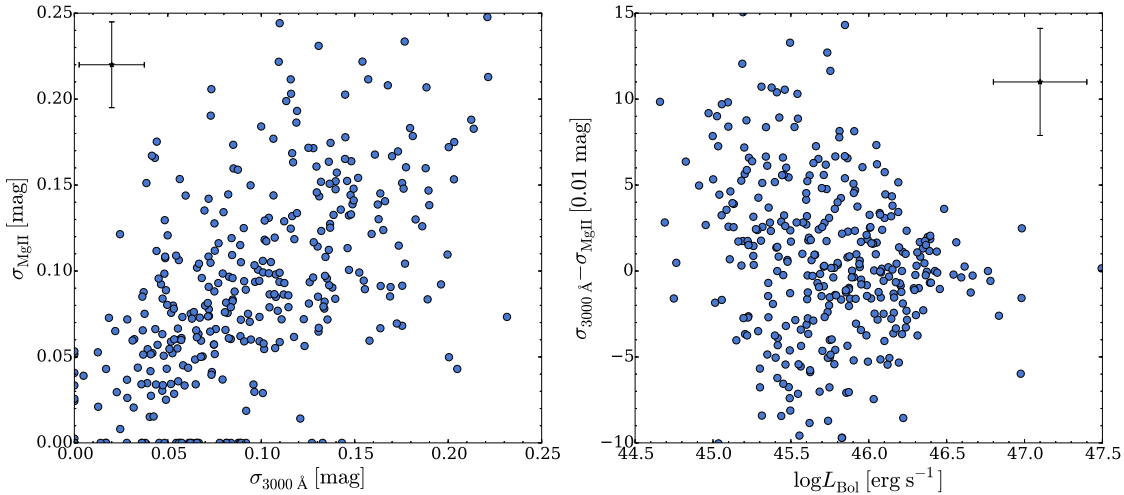}
\caption{Left: The light-curve variability of \MgII \ against 
that of the $3000\ \rm{\AA}$ continuum. The light-curve variability 
of the two components are correlated. Right: The difference 
between the light-curve variability of the $3000\ \rm{\AA}$ continuum 
and that of \MgII . The difference is anti-correlated with quasar 
luminosity. }
\label{fig:mgii-l3k-lcvar}
\end{figure*}

We compare the light-curve variability of \MgII \ with that of 
the $3000\ \rm{\AA}$ continuum in the left panel of 
Figure~\ref{fig:mgii-l3k-lcvar}. The Spearman correlation test 
again reveals $\rho=0.62$ and the $p$ value is $6\times10^{-39}$, 
i.e., our data favor a significant positive correlation between 
the light curve variability of \MgII \ and that of the $3000\ 
\rm{\AA}$ continuum. This correlation indicates that the variability 
in the light curve of \MgII \ and that of the $3000\ \rm{\AA}$ 
continuum are connected. The simplest explanation is that \MgII \ 
varies in response to the $3000\ \rm{\AA}$ continuum \citep[see 
the cross correlation analysis of][]{shen15c}. Furthermore, the 
scatter in the correlation (for instance, a few sources show 
significant variability in the $3000\ \rm{\AA}$ continuum but the 
variability of \MgII \ is consistent with $0$) suggests that 
the response process may not be uniform in all quasars, e.g., 
the response of the emission line to the continuum and/or the 
structure of the BLR may differ in different quasars. 

In the right panel of Figure~\ref{fig:mgii-l3k-lcvar}, we plot the 
difference between the light curve variability of the $3000\ \rm{\AA}$ 
continuum and that of \MgII \ versus quasar luminosity. Most 
quasars vary slightly more in the $3000\ \rm{\AA}$ continuum than 
in the \MgII \ line; the median $\sigma_{lc,3000}-\sigma_{lc,\rm{MgII}}$ 
is $0.003\ \rm{mag}$. A Spearman rank correlation test between the 
difference in the $3000\ \rm{\AA}$ continuum and \MgII \ variability 
and $L_{\rm Bol}$ indicates that the difference is anti-correlated 
with quasar luminosity, with $\rho=-0.2$ and a $p$ value of $10^{-4}$; 
as quasar luminosity increases, the variability of the $3000\ \rm{\AA}$ 
continuum decreases more rapidly than the \MgII \ variability. 

\subsubsection{\Hbeta}
We then present the light-curve variability of \Hbeta . Since our 
final sample of \Hbeta \ consists of only $41$ quasars, we interpret 
our data with caution. 

\begin{figure}
\epsscale{1.15}
\plotone{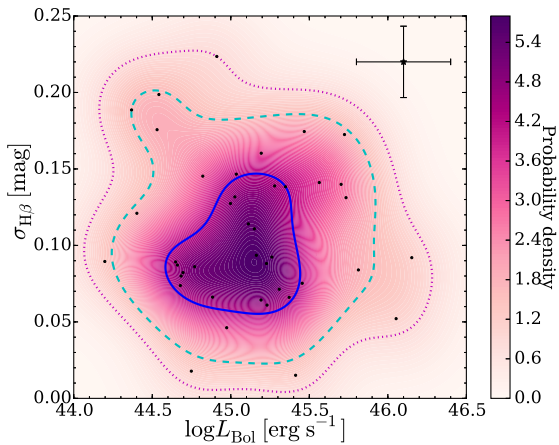}
\caption{The light-curve variability of \Hbeta . Each point 
represents a quasar. The solid, dashed, and dotted lines 
represent $1\sigma$, $2\sigma$ and $3\sigma$ bounds of the 
quasar number distribution. The color indicates the probability 
density of each point. The 
light-curve variability of \Hbeta \ does not demonstrably 
depend on quasar luminosity which may due to the fact that 
there are only $41$ quasars in the sample.}
\label{fig:hbeta-lcvar}
\end{figure}

Figure~\ref{fig:hbeta-lcvar} shows the light-curve variability 
of \Hbeta \ as a function of quasar luminosity. 
The Spearman correlation test suggests that there is no 
significant correlation between the variability and quasar 
luminosity ($\rho= -0.09$ with the $p$ value of $0.58$). This 
result may be due to the small size of our sample. We test 
such small-sample effects by randomly selecting 
$41$ quasars from the \MgII \ clean sample and testing the 
correlation between the \MgII \ light-curve variability and quasar 
luminosity. We repeated this simulation $10^4$ times and 
found that $\sim 15\%$ of the time, the anti-correlation between 
\MgII \ variability and quasar luminosity of the limited sample 
is as weak as or weaker than the Spearman correlation 
test for \Hbeta \ variability with quasar luminosity. 
From this simulation, we conclude that the lack of the correlation 
between the variability of \Hbeta \ and quasar luminosity is 
plausibly caused by the small sample size. We then 
tested the correlation between the light-curve variability of \Hbeta \ 
and the Eddington ratio ($\rho=-0.33$, and the $p$ value is 
$0.035$) and found the anti-correlation is not statistically significant. 
We performed the same simulation and found that the lack of 
correlation is again plausibly due to the small sample size. 

\begin{figure*}
\epsscale{1.15}
\plotone{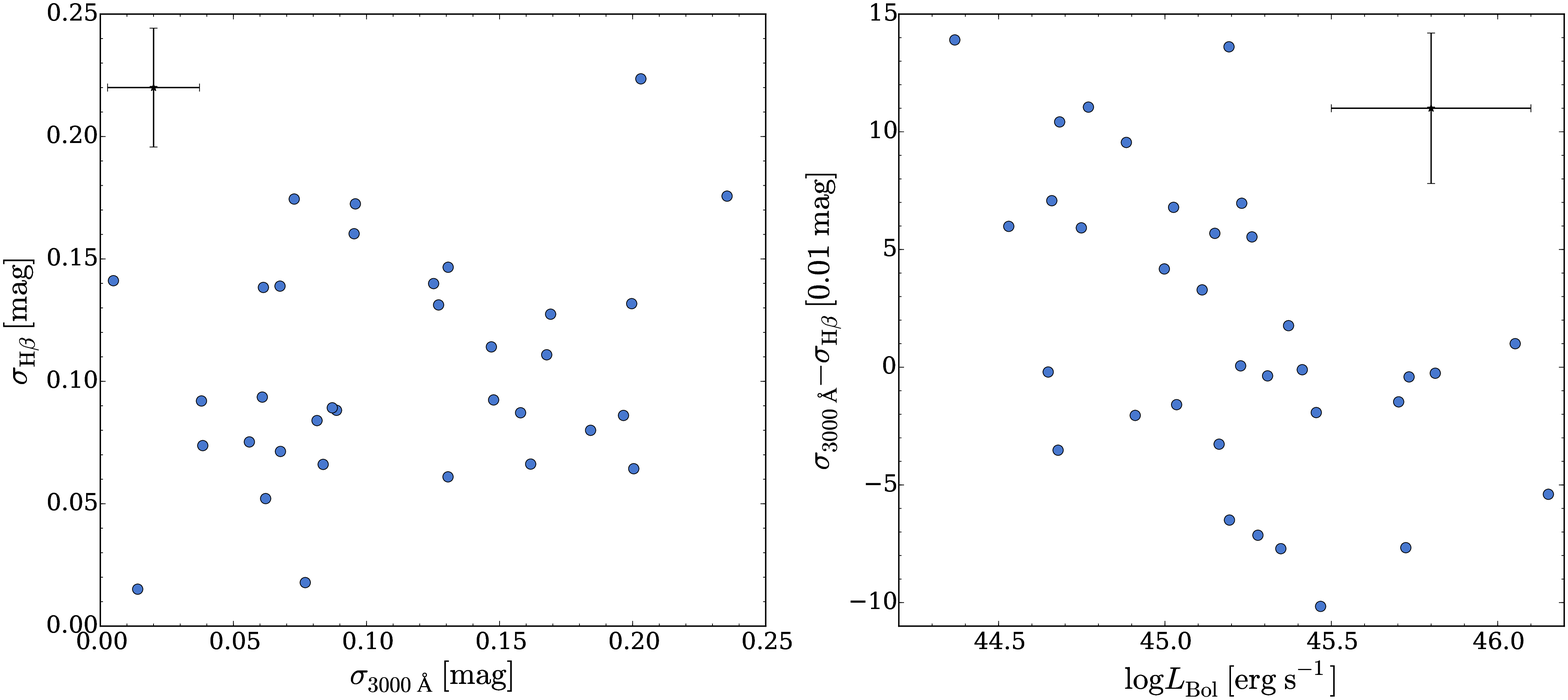}
\caption{Left: The light-curve variability of \Hbeta \ against 
that of the $3000\ \rm{\AA}$ continuum. There is no significant 
correlation between the variability of the two components. 
Right: The difference between the light-curve variability of the 
$3000\ \rm{\AA}$ continuum and that of \Hbeta . The difference 
is anti-correlated with quasar luminosity.}
\label{fig:hbeta-l3k-lcvar}
\end{figure*}

We then compare the light-curve variability of \Hbeta \ with 
that of the $3000\ \rm{\AA}$ continuum. We prefer the $3000\ 
\rm{\AA}$ continuum instead of the $5100\ \rm{\AA}$ continuum 
because the galaxy contamination to the $5100\ \rm{\AA}$ 
continuum is more significant (and galaxy contamination biases 
the observed variability, see Section~\ref{sec:sf5k}). 
We can only compare quasars for which \Hbeta \ 
and the $3000\ \rm{\AA}$ continuum were both observed in the 
BOSS spectrum ($35$ quasars). The left panel of 
Figure~\ref{fig:hbeta-l3k-lcvar} presents the light curve 
variability of \Hbeta \ as a function of that of the $3000\ 
\rm{\AA}$ continuum. The Spearman correlation test indicates 
that there is no significant correlation between the variability 
of \Hbeta \ and that of the $3000\ \rm{\AA}$ continuum ($\rho=0.28$, 
and the $p$ value is $0.1$). Once again, we test if the lack 
of correlation is caused by small sample size by testing for 
correlations in random subsets of $35$ quasars from the \MgII \ 
clean sample. In $10^4$ simulations, the \MgII \ and continuum 
variability are uncorrelated $\sim 0.5\%$ of the time. From this we 
conclude that the lack of correlation between \Hbeta \ and continuum 
variability is not solely due to selection effects. Given the 
fact that previous reverberation mapping work reveals that \Hbeta \ 
does respond to continuum variability, our results indicate that 
\Hbeta \ and \MgII \ variability relate to continuum variability in 
different ways. For example, the intrinsic correlation between 
the \Hbeta \ variability and continuum variability might be slightly 
weaker. It is also possible that the transfer function of  \Hbeta \ 
increases significantly with quasar luminosity, thus making the 
\Hbeta \ variability behave opposite to continuum variability as 
a function of quasar luminosity. The different relationships of 
\Hbeta \ and \MgII \ variability with quasar luminosity in turn 
probe the differences between the BLR gas responsible for each 
line (see Section~\ref{sec:dis3}). 

In the right panel of Figure~\ref{fig:hbeta-l3k-lcvar}, we plot 
the difference between the variability of the $3000\ \rm{\AA}$ 
continuum and that of \Hbeta \ as a function of quasar luminosity. 
We again tested the correlation between the difference of the 
variability and quasar luminosity using the Spearman correlation 
test. We found that there is an anti-correlation between the 
differences of the variability and quasar luminosity ($\rho=-0.56$, 
and the $p$ value is $4\times10^{-4}$). This correlation makes 
sense given the previously-found anti-correlation of the $3000\ 
\rm{\AA}$ continuum variability with quasar luminosity, even 
with the lack of correlation of \Hbeta \ variability with quasar 
luminosity. 

\begin{figure*}
\epsscale{1.15}
\plotone{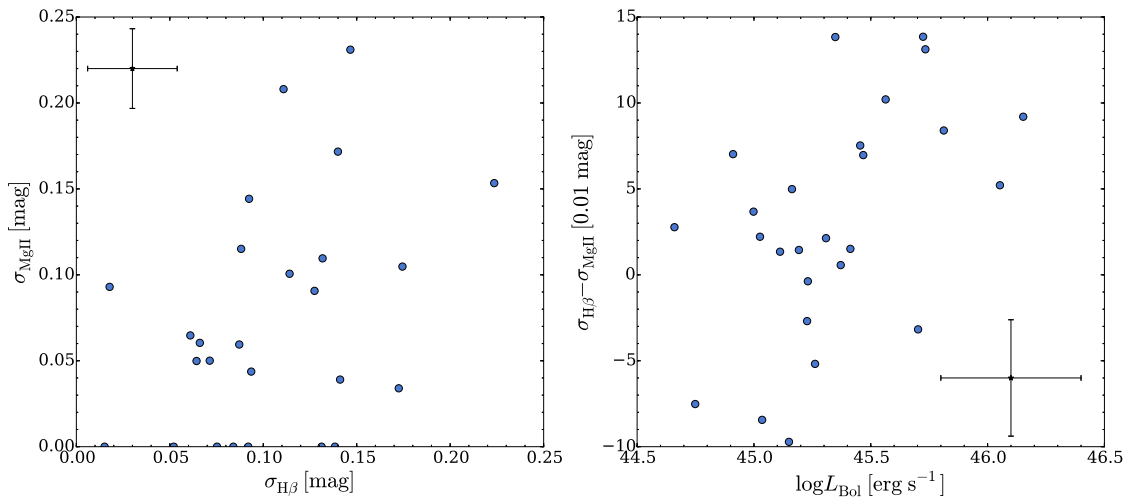}
\caption{Left: The light-curve variability of \Hbeta \ against 
that of \MgII . There is no significant correlation between the 
variability of the two components. Right: The difference 
between the light-curve variability of \Hbeta \ and that of \MgII . 
The difference is weakly correlated with quasar luminosity.}
\label{fig:hbeta-mgii-lcvar}
\end{figure*}

We performed similar comparisons between \Hbeta \ and \MgII . 
Our results are presented in Figure~\ref{fig:hbeta-mgii-lcvar}. 
Again, we can only consider quasars for which \MgII \ and \Hbeta \ 
were both observed in the BOSS spectrum ($26$ quasars). The 
Spearman correlation test suggests that there is no strong correlation 
between the variability of \Hbeta \ and that of \MgII \ ($\rho=0.36$, 
and the $p$ value is $0.07$), and on average, the variability observed 
in \MgII \ is less than observed for \Hbeta . Note that there are seven 
out of $26$ sources that show significant variability in the light curve 
of \Hbeta \ but not in that of \MgII . This is probably due to 
the fact that, as \MgII \ is on average less variable than \Hbeta , 
the observed light-curve variability of \MgII \ is more likely to be 
dominated by measurement and spectrophotometric errors. The lack 
of correlation between \Hbeta \ and \MgII \ variaiblity is likely due to 
the small sample size and/or the differences between the BLR gas 
that produce \Hbeta \ and \MgII \ (see Section~\ref{sec:dis3}). 

There might be a weak correlation between the difference of the 
variability of \Hbeta \ and that of \MgII \ and quasar luminosity 
as revealed by the Spearman test ($\rho=0.51$, and the $p$ 
value is $0.008$) although the Kendall test suggests that we cannot 
rule out the no-correlation hypothesis (the $p$ value is $0.02$). 
This correlation, if indeed exist, can also be explained 
by the anti-correlation between the variability of \MgII \ and 
quasar luminosity (which holds even we only consider these $26$ 
quasars), even with the lack of correlation of \Hbeta \ 
variability with quasar luminosity. 

\section{Observed Distribution of $\Delta m$}
\label{sec:obsvar}
Previous studies have revealed that, in general, intrinsic 
quasar variability is not consistent with white noise but 
is actually a red-noise process where quasars are more 
variable on longer timescales \citep[e.g.,][hereafter 
M10]{mac10}. The basic light-curve variability we presented 
in Section~\ref{sec:lcv} is averaged over different timescales. 

\begin{figure}
\epsscale{1.2}
\plotone{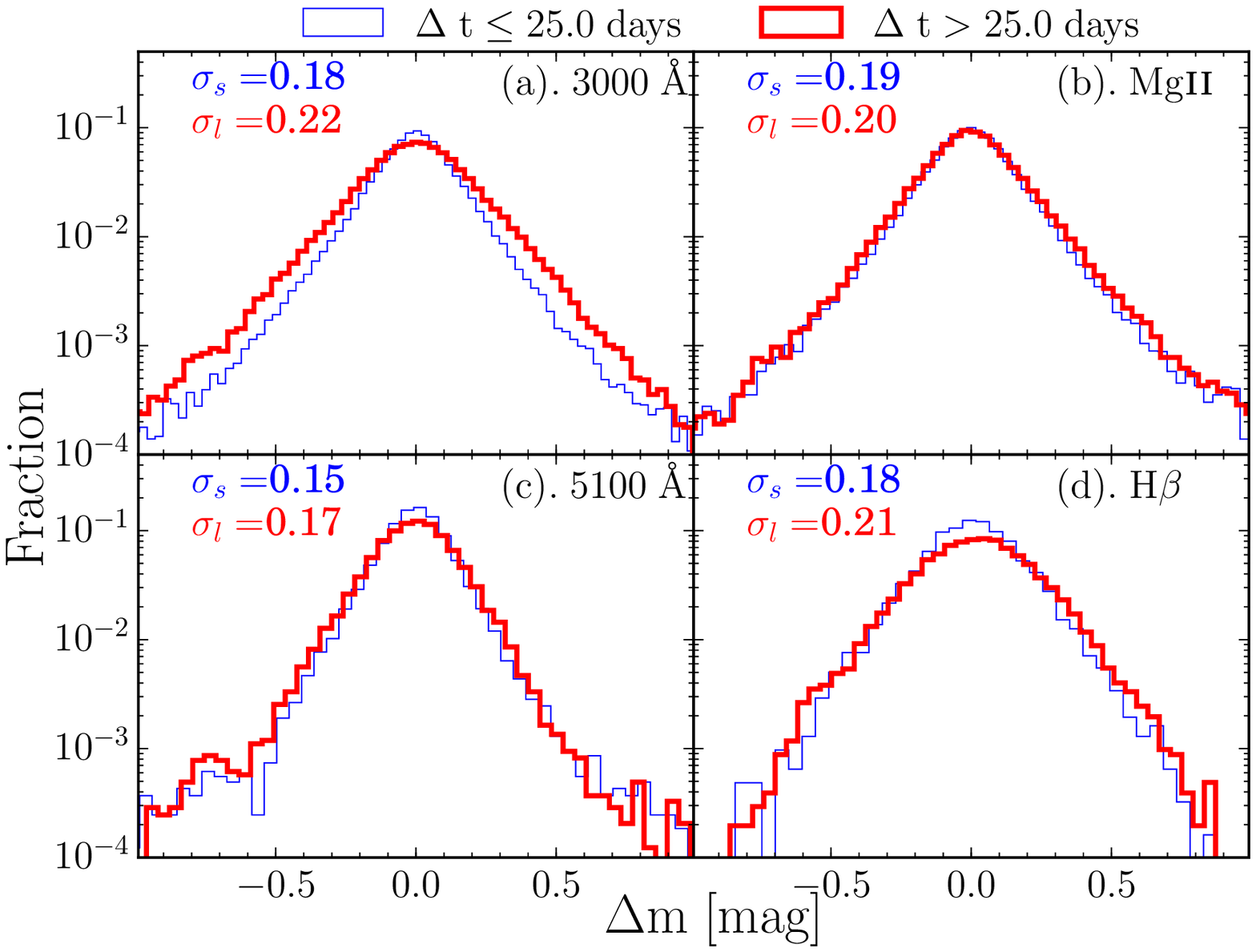}
\caption{The histograms of observed $\Delta m$ of the SDSS-RM 
sources. In this figure, $\Delta m=-2.5\log (f_2/f_1)$, where 
$f_1$ and $f_2$ are two measurements of flux separated by 
rest-frame $\Delta t$. The blue and red lines represent the 
variability of $\Delta t\leq 25\ \rm{days}$ (the median $\Delta 
t\sim 12\ \rm{days}$ for $L_{3000}$ and \MgII \ and $\sim 14\ 
\rm{days}$ for $L_{5100}$ and \Hbeta ) and $\Delta t>25\ 
\rm{days}$ (the median $\Delta t\sim 40\ \rm{days}$ for $L_{3000}$ 
and \MgII \ and $\sim 50\ \rm{days}$ for $L_{5100}$ and \Hbeta ), 
respectively. $\sigma_s$ and $\sigma_l$ in each panel represent 
the standard deviation of the short and long timescale distributions, 
respectively. There are exponential tails in the distribution of 
$\Delta m$. This is likely due to the superposition of the intrinsic 
quasar variability (which is a superposition of many Gaussian 
distributions), measurement errors, and spectrophotometric errors. 
}
\label{fig:var-rm}
\end{figure}

Before we calculate the quasar variability as a function of 
rest-frame $\Delta t$ (i.e., the structure function), we first 
present the distribution of luminosity variability for two 
different $\Delta t$ bins. We calculated $\Delta m=-2.5\log 
(f_2/f_1)$ for all luminosity pairs separated by $\Delta t$. 
The histograms of $\Delta m$ for the continuum and broad 
emission lines are presented in Figure~\ref{fig:var-rm}. Note 
that the observed histograms are a superposition of the 
intrinsic variability, measurement errors, and spectrophotometric 
errors (see Appendix~\ref{sec:sperr}). The distribution of 
$\Delta m$ is not Gaussian for any component. Actually, 
the distributions are better described by the Laplace (i.e., 
double-exponential) distribution. This can be explained either 
by the fact that the spectrophotometric errors are not Gaussian 
(see Appendix~\ref{sec:sperr}) or, as illustrated by \cite{mac12} 
(their Section 3.2.2), as natural results of a superposition of 
many Gaussian distributions. 

\begin{figure}
\epsscale{1.2}
\plotone{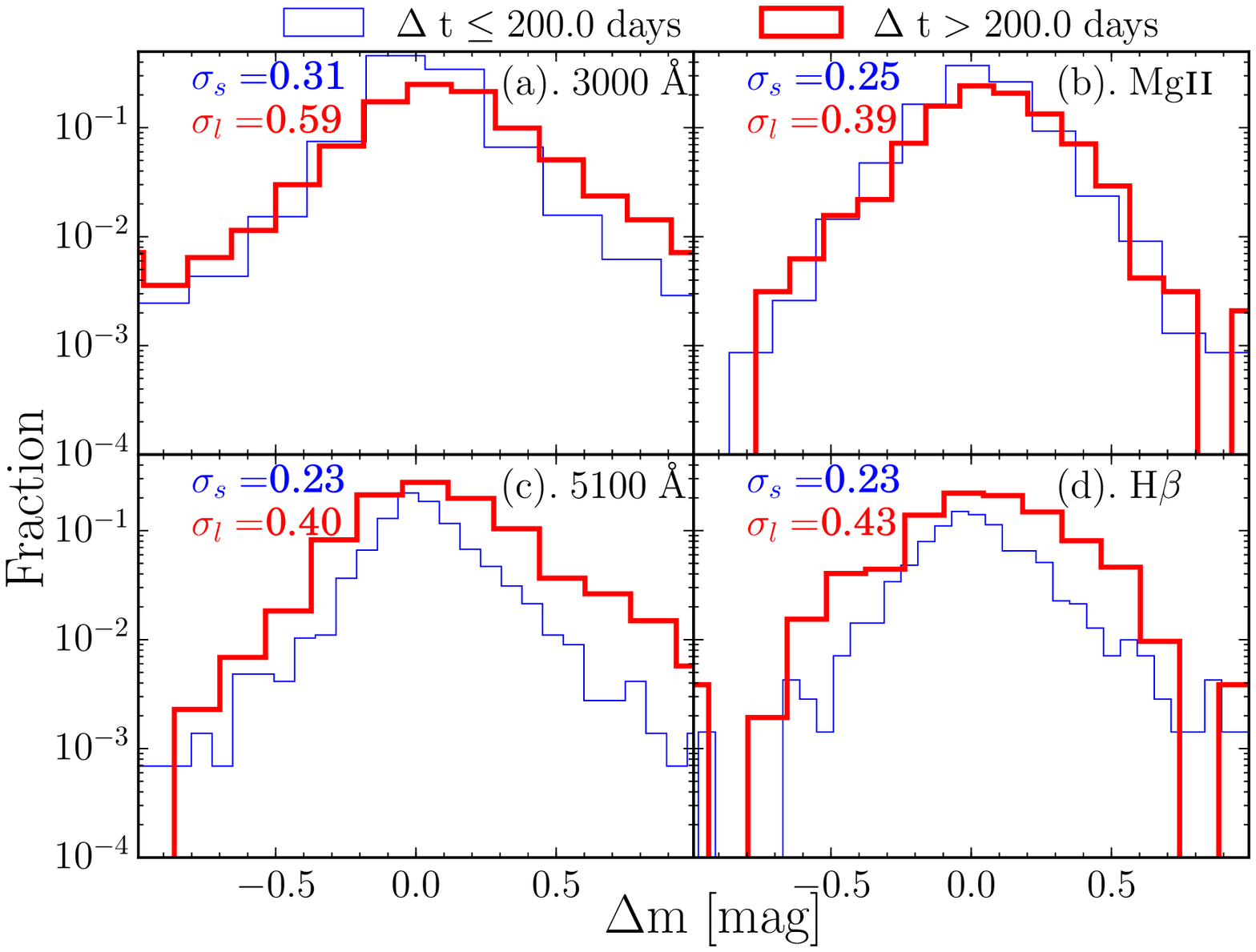}
\caption{The histograms of $\Delta m$ of the SDSS-I/II sources. The blue 
and red lines represent the variability of $\Delta t\leq 200\ \rm{days}$ 
(the median $\Delta t\sim 20\ \rm{days}$) and $\Delta t>200\ \rm{days}$ 
(the median $\Delta t\sim 300\ \rm{days}$), respectively. There are 
exponential tails in the distribution of $\Delta m$. Also, the variability 
increases with $\Delta t$.}
\label{fig:var-12}
\end{figure}

The distributions of $\Delta m$ for SDSS-I/II sources are shown in 
Figure~\ref{fig:var-12}. The SDSS-I/II distributions are less 
well-sampled in $\Delta m$ because each quasar only has two epochs, 
compared to the $29$ epochs of the SDSS-RM sources. In addition, the 
increase in variability with $\Delta t$ is clearer in Figure~\ref{fig:var-12} 
because of the larger range of $\Delta t$ in the SDSS-I/II data (which 
span tens to thousands of days). Otherwise, the SDSS-I/II variability 
distribution is similar to our SDSS-RM sources, with similar exponential 
tails. 

\section{The Structure Function: Variability 
as a Function of Time Separation}
\label{sec:sfall}
In this section, we introduce the structure function, a tool to study 
quasar variability as a function of rest-frame time separation 
($\Delta t$) between observations.

\subsection{Definitions}
\label{sec:def}
The structure function, $\mathrm{SF}(\Delta t)$, measures the 
statistical dispersion of observations (e.g., luminosity or flux) 
separated by time intervals, $\Delta t$. Many non-parametric 
statistics can be used as estimators of the statistical dispersion 
of the distribution of $\Delta m$, e.g., the interquartile range 
(IQR), the median absolute deviation (MAD), the average absolute 
deviation (AAD), and the standard deviation. The IQR and MAD 
statistics tend to be more robust against outliers or tails in the 
distribution than the AAD or standard deviation statistics: see 
Appendix~\ref{sec:sfest} for more details. In this work, we adopt 
the IQR estimator as it has been adopted in previous work. In order 
to account for the presence of measurement and spectrophotometric 
errors in our data, we estimate the intrinsic statistical dispersion 
following \cite{mac12}: 
\begin{equation}
\label{eq:iqr}
\mathrm{SF_{IQR}}(\Delta t) = \sqrt{(0.74\mathrm{IQR}(\Delta m))^2 - 
\widetilde{\sigma^2_{\rm{e}}}} \,
\end{equation}
where $\mathrm{IQR}(\Delta m)$ is the $25\%-75\%$ interquartile range 
of $\Delta m$ and $\sigma_{\rm{e}}$ is the total uncertainty of $\Delta m$ 
(i.e., the summation of measurement error and spectrophotometric error 
in quadrature). The constant $0.74$ normalizes the IQR to be equivalent 
to the standard deviation of a Gaussian distribution; the IQR is smaller 
than the standard deviation for a Laplace distribution. 
Appendix~\ref{sec:sfest} also includes discussion of alternative structure 
function estimators: AAD, MAD, and a maximum-likelihood estimator of 
standard deviation. 

The structure function can also be used to characterize the statistical 
dispersion of $\Delta m$ for a sample of many quasars with the same 
(or close) $\Delta t$. That is, the ensemble structure function\footnote{
Limited time sampling can cause spurious features in the observed 
structure function for individual objects \citep{emm10}, but by averaging 
many quasar light-curves the ensemble structure function produces a 
better-sampled rest-frame time coverage. In Sections~\ref{sec:sfcont} 
\&~\ref{sec:sfeml}, we also perform simulations 
to account for possible sampling issues.} is an average of the structure 
functions of many sources \citep[as demonstrated by][individual and ensemble 
structure functions result in the same statistical properties]{mac08}. All these 
structure function estimators involve square roots, and therefore must not be 
negative. If measurement and spectrophotometric errors are significantly 
larger than the intrinsic variability of quasars and/or the measurement 
errors are underestimated by a factor of a few, the ensemble structure 
function would be strongly biased. Our sample's $S/N>10$ requirement 
is designed to minimize this bias.

\subsection{The Damped Random-Walk Model}
\label{sec:drw}
As pointed out by previous work \citep[e.g.,][]{kelly09,mac12,zu13}, 
the light curves of quasar continuum emission can be well described 
by the DRW model. The DRW model describes a random process that 
is characterized by the following covariance matrix:
\begin{equation}
C(x(t_i), x(t_j)) = \mathrm{SF}_{\infty}^2\exp(-\Delta t/\tau)/2
\end{equation}
where $\Delta t=\left| t_i-t_j \right|$. $\mathrm{SF}_{\infty}$ and 
$\tau$ are the asymptotic driving amplitude and the damping timescale. 
The structure function is given by 
\begin{equation}
\label{eq:sfdrw}
{\mathrm SF}(\Delta t|\tau,{\rm SF}_{\infty}) = {\rm SF}_{\infty} 
\sqrt{1-\exp(-\Delta t/\tau)}
\end{equation}
At $\Delta t \ll \tau$ (with $\tau$ typically on the order of hundreds of 
days), Eq.~\ref{eq:sfdrw} can be 
simplified as, 
\begin{equation}
\label{eq:short}
\mathrm{SF}(\Delta t)=\hat{\sigma}\sqrt{\Delta t}
\end{equation}
where $\hat{\sigma}=\rm{SF}_{\infty}/\sqrt{\tau}$, and the structure 
function at $\Delta t=100\ \rm{days}$, $\rm{SF}_{100}=10\hat{\sigma}$. 
The ensemble structure function for given luminosity, $L$, bin is given 
by 
\begin{equation}
\mathrm{SF_{EN}}(\Delta t) = \int_{\Delta t_{\mathrm{min}}}^{\Delta 
t_{\mathrm{max}}} \int \int d\Delta t_0 d\tau d SF_{\infty}\ SF(\Delta 
t_0|\tau, SF_{\infty})
\end{equation} 
For a quasar, $\tau$ and $\rm{SF}_{\infty}$ (or $\hat{\sigma}$) are 
determined by $M_{\rm BH}$, $L$, and wavelength (e.g., M10). For 
a set of quasars binned in a narrow range of $M_{\rm BH}$, $L$, 
and wavelength, $\tau$ and $\rm{SF}_{\infty}$ (or $\hat{\sigma}$, 
$\rm{SF}_{100}$) are constant and the integrals above reduce to a 
simple summation.

\section{The Structure Functions of 
Continuum emission}
\label{sec:sfcont}
The main purpose of this work is to investigate the variability 
of broad emission lines and their connection to the variability 
of continuum emission. Therefore, we first present the 
structure function of the continuum emission.

\subsection{The $3000$ \rm{$\rm{\AA}$} Continuum}
\label{sec:sf3k}
\begin{figure*}
\epsscale{1.3}
\plotone{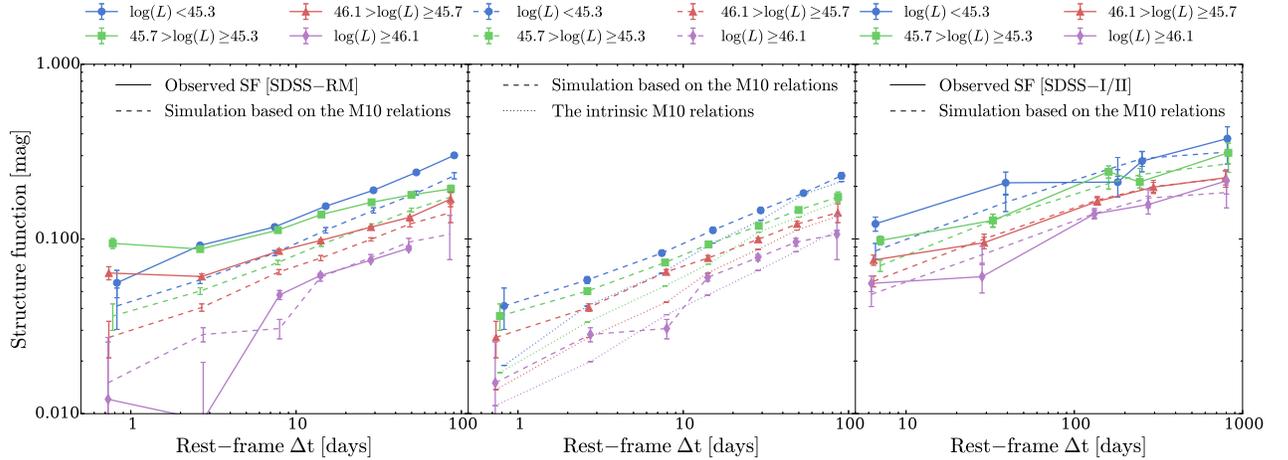}
\caption{Left panel: The structure function (solid lines) of 
the $3000\ \rm{\AA}$ continuum of the SDSS-RM sources for 
each luminosity bin. The dashed lines represent 
our simulation results based on the M10 relations ``observed'' 
like our data (see the text in Section~\ref{sec:sf3k}). The 
structure functions increase with $\Delta t$ and decrease 
with $L_{\rm{Bol}}$. The shapes of the observed structure 
functions (on timescales $\Delta t>5\ \rm{days}$) are 
consistent with the DRW model and the M10 relations (the 
difference in slope $\beta \lesssim 0.1$). The 
M10 relations slightly under-predict (by a factor of $1.3$) 
the variability amplitude. 
Middle Panel: The dotted lines represent the structure functions 
from the M10 relations without considering the observational bias. 
On short timescales, the measurement and spectrophotometric 
uncertainties of our data dominate over the intrinsic quasar 
variability, and the structure function estimations are biased. 
The bias is not significant on longer timescales where the 
intrinsic quasar variability is larger. 
Right panel: The structure functions (solid lines) of the $3000\ 
\rm{\AA}$ continuum for SDSS-I/II sources. The dashed lines 
represent the structure functions predicted by the M10 relations. 
The M10 relations can roughly reproduce the observed structure 
functions. According to the M10 relations, for the four luminosity 
bins, the characteristic timescale $\tau\sim 200\ \rm{days}$. In 
this figure and the rest of the figures, the ($1\sigma$) uncertainties 
of the structure functions are estimated via bootstrapping, and $L$ 
represents $L_{\rm{Bol}}$.}
\label{fig:sf3k}
\end{figure*}

In this section, we present the structure function of the $3000\ 
\mathrm{\AA}$ continuum for the SDSS-RM sources. We first 
divided our sample into $4$ sub-samples by $L_{\mathrm{Bol}}$: 
$L_{\mathrm{Bol}}<10^{45.3}\ \mathrm{erg\ s^{-1}}$; $10^{45.3}\ 
\mathrm{erg\ s^{-1}} \leq L_{\mathrm{Bol}} <10^{45.7}\ \mathrm{erg\ 
s^{-1}}$; $10^{45.7}\ \mathrm{erg\ s^{-1}} \leq L_{\mathrm{Bol}} 
<10^{46.1}\ \mathrm{erg\ s^{-1}}$; $L_{\mathrm{Bol}}> 10^{46.1}\ 
\mathrm{erg\ s^{-1}}$. The luminosity bins are constructed 
to have (roughly) equal numbers of quasars in each. We calculated 
the ensemble structure function in each bin using the method 
described in Section~\ref{sec:def} (subtracting both 
measurement and spectrophotometric errors). The binned IQR 
structure functions for the $3000\ \rm{\AA}$ continuum of the 
SDSS-RM quasars are shown by the solid lines in the left and 
center panels of Figure~\ref{fig:sf3k}. 

Using a Stripe 82 quasar sample, M10 found that the DRW 
parameters are correlated with quasar properties in the 
following way,
\begin{equation}
\label{eq:m10}
\log q = Q_1 + Q_2\log(\frac{\lambda_{\mathrm{rf}}}{4000 \ 
\mathrm{\AA}}) + Q_3(M_i+23) + Q_4\log(\frac{M_{\mathrm{BH}}}{10^9\ 
M_{\odot}}) \\,
\end{equation}
where for $q=SF_{\infty}$, $Q_1=-0.51$, $Q_2=-0.479$, 
$Q_3=0.131$, and $Q_4=0.18$; for $q=\tau$, $Q_1=2.4$, 
$Q_2=0.17$, $Q_3=0.03$, and $Q_4=0.21$ (here $\tau$ is 
in units of days). The uncertainty of each coefficient 
can be found in M10. The intrinsic scatters of the M10 
relations are presented in \cite{mac12}. $M_i$ is the 
K-corrected rest-frame $i$-band absolute magnitude. 

To compare our results with the M10 relations (i.e., 
Eq.~\ref{eq:m10}), we performed a simulation based 
on the M10 relations (hereafter ``DRW simulation''). 
The simulation procedures are similar to those of 
\cite{mac12}: 
\begin{enumerate}
\item We calculate a model structure function for each 
quasar using its $M_{\rm BH}$ and $L_{\rm Bol}$ 
(translated into rest-frame $M_i$ assuming the $i$-band 
bolometric correction factor is $12$), following Eq.~\ref{eq:m10}. 
During the calculation, $M_{\rm BH}$, $L_{\rm Bol}$, 
and the coefficients in Eq.~\ref{eq:m10} are perturbed 
by Gaussian noise according to their uncertainties. The 
predicted $\hat{\sigma}$ (i.e., $SF_{\infty}/ \sqrt{\tau}$) 
and $\tau$ (perturbed by Gaussian noise according to 
the intrinsic scatter in Eq.~\ref{eq:m10}) are translated 
into a structure function value at each $\Delta t$ using 
Eq.~\ref{eq:sfdrw}. 

\item In each bin of quasar luminosity at each $\Delta t$, 
we randomly generated a distribution of $\Delta m$ using 
the model structure function for each quasar in that bin. 
We then add the measurement and spectrophotometric 
errors (following Laplace, not Gaussian, distributions; see 
Appendix~\ref{sec:sfstar}) to create the ``observed'' 
variability of the model. 

\item We calculated the structure function (using the IQR 
estimator) from the model distribution of $\Delta m$, 
subtracting the median errors in quadrature in each bin 
(see Section~\ref{sec:def}).
\end{enumerate}
By doing this, the simulation includes the same numerical 
and/or time-sampling issues as the structure function 
calculated from the real data. We repeated this simulation 
$256$ times to obtain a simulated structure function 
for each luminosity bin.

Our results are shown in the left panel of Figure~\ref{fig:sf3k} 
(for other structure-function estimators, see 
Appendix~\ref{sec:sfest}). It is evident that the variability 
amplitude decreases with quasar luminosity and increases 
with $\Delta t$. Also, the shape of the observed structure 
function in each luminosity bin can be reproduced well by 
a simulated DRW model ``observed'' like the data (i.e., with 
the same time sampling and measurement/spectrophotometric 
errors). To verify this, we performed a statistical hypothesis 
test. 

The null hypothesis of our test is the following: for 
each luminosity bin, the observed structure function and the 
simulated structure function share the same shape but with 
different normalization factors (in $\log \mathrm{SF}(\Delta 
t)-\log \Delta t$ space). We calculated the difference 
between the observed $\log \mathrm{SF}(\Delta t)$ and the 
simulated $\log \mathrm{SF}(\Delta t)$ and their uncertainties 
for each luminosity bin. Note that we only considered data 
points with $\Delta t> 5\ \rm{days}$ (see Section~\ref{sec:dis1}). 
If our null hypothesis is true, then the difference we obtained 
should be, statistically, a constant (i.e., independent of 
$\Delta t$). We adopted the chi-squared test to assess this 
hypothesis. For three of the four luminosity bins considered, 
our data fail to reject the null hypothesis that the two 
structure functions share the same shape (i.e., the $p$ value 
of the chi-squared test is $\gg 0.01$). The simulated and observed 
structure functions have different shapes only in the $10^{45.3}\ 
\mathrm{erg\ s^{-1}} \leq L_{\mathrm{Bol}} <10^{45.7}\ \mathrm{erg\ 
s^{-1}}$ bin (although this happens only $25\%$ of the time), 
but differ in slope by only $\Delta \beta \simeq 0.1$ (parameterizing 
each by $\mathrm{SF}(\Delta t)\sim \Delta t^{\beta}$). 

To characterize the sensitivity of our statistical test, we 
simulated two pairs of structure functions (both follow 
$\mathrm{SF}(\Delta t)\sim \Delta t^{\beta}$) with different 
slopes $\beta$ and with the same S/N as our observed structure 
functions and our ``DRW simulation'' structure functions. We 
then applied our statistical test to each pair of simulated 
structure functions and calculated the $p$ value of the null 
hypothesis that the shapes are the same. We repeated this 
simulation $10^4$ times and found that in $80\%$ of the 
simulations (which is the most widely adopted value in 
statistical power analysis), our null hypothesis can be 
rejected if the difference in $\beta$ is $>0.1$ ($> 0.2$, 
for the highest luminosity bin): the difference in 
slopes is $\lesssim 0.1$.\footnote{Throughout this work, 
the maximum ``allowed'' difference in $\beta$ is estimated in 
this way. } Therefore, we 
conclude that, for each luminosity bin, the observed structure 
function and the structure function from our ``DRW simulation'' 
share the same shape (the difference in slope $\beta \lesssim 
0.1$). That is, as revealed by previous work 
\citep[e.g.,][]{kelly09, mac10, mac12}, the variability 
of the quasar continuum emission can be described by the DRW 
model. 

As for the variability amplitude, we find that the M10 relations 
under-predict the amplitude (by a factor of $\sim 1.3$) except 
for the highest luminosity bin (i.e., $\log L_{\rm{Bol}}>46.1$). 
This indicates that while the M10 relations are accurate 
for high-luminosity quasars, a small revision is required 
to reproduce the variability of lower luminosity quasars. 
Note that the quasar sample used in M10 consists of luminous 
quasars ($M_i<-23$) while a significant fraction of quasars 
in our sample are less luminous ($M_i>-23$). That is, our 
lower-luminosity quasars require extrapolation of the M10 
relations, and this extrapolation seems to be inaccurate 
by a small factor. 

In the middle panel of Figure~\ref{fig:sf3k}, we illustrate 
the importance of ``observing'' the model structure function, 
comparing the ``observed'' model structure function to the 
basic model without bias treatment. On long timescales 
($\sim 20-100$ days), the two versions of the structure 
functions are similar. On short timescales ($<20$ days), 
the two versions of structure function are different and the 
differences increase with decreasing $\Delta t$. This is 
because, on short timescales, the uncertainties of our data 
are comparable to or even dominate over the intrinsic variability, 
and the bias of the observed structure function is significant. 
In this case, if we add uncertainties and then subtract them 
via Eq~\ref{eq:iqr}, the model becomes biased toward higher 
amplitude variability.

We then studied the structure function of the $3000\ \rm{\AA}$ 
continuum on longer ($\Delta t\sim 1000\ \rm{days}$) 
timescales using the ancillary SDSS-I/II data. Similar to 
our SDSS-RM sources, we divided the sample into four bins 
of $L_{\rm{Bol}}$ and calculated the structure function in 
each bin. We then test whether the M10 relations can 
effectively describe our data by performing the 
``DRW simulation'' for SDSS-I/II sources. 

The right panel of Figure~\ref{fig:sf3k} shows our results. 
Similar to the SDSS-RM sources, the structure function of 
the $3000\ \rm{\AA}$ continuum for SDSS-I/II quasars also 
increases with $\Delta t$ and decreases with $L_{\rm{Bol}}$. 
On timescales of $\sim 10-100\ \rm{days}$, the SDSS-RM and 
SDSS-I/II quasars have similar variability amplitudes and 
the M10 relations again under-predict the variability amplitude 
by a factor of $1.2$ for the three low-luminosity bins. The 
M10 relations successfully reproduce the shape of the observed 
structure function (the same test reveals a $p$ value $\gg 0.01$, 
and the difference in slope $\beta$ is $\lesssim 0.15$) for 
each luminosity bin. In addition, using the M10 relation 
(Eq.~\ref{eq:m10}), we calculate the ensemble characteristic 
timescale $\tau$ and find 
that $\tau\sim 199.5^{+116.7}_{-73.6}$ days\footnote{Throughout 
this work, the reported uncertainties are $1\sigma$.} (where the 
error comes from the 
propagated uncertainties in $M_{\rm BH}$, $L_{\rm Bol}$, 
and the coefficients in Eq.~\ref{eq:m10} and the intrinsic scatter 
for $\tau$ in Eq.~\ref{eq:m10}).

\subsection{The $5100$ \rm{$\rm{\AA}$} \it{Continuum}}
\label{sec:sf5k}
\begin{figure*}
\epsscale{1.2}
\plotone{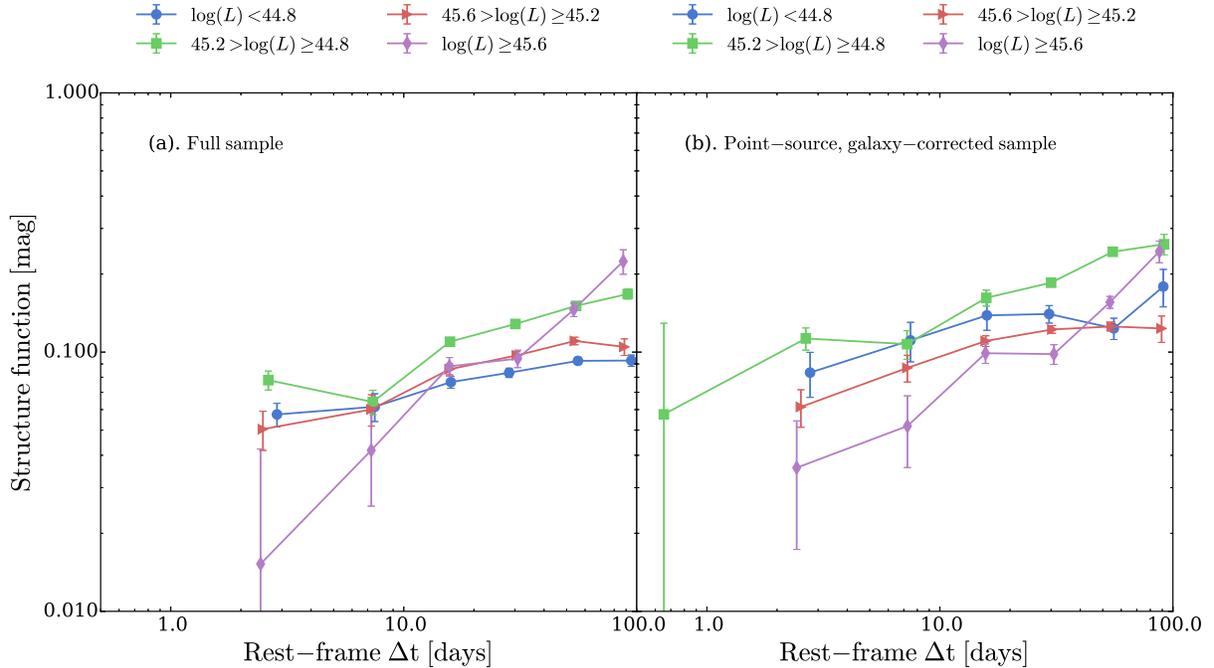}
\caption{Left panel: The structure functions of the $5100\ \rm{\AA}$ 
continuum for the SDSS-RM sample. The structure functions roughly 
increase with $\Delta t$ but are not monotonic functions of $L_{\rm{Bol}}$. 
This is likely due to the contamination by galaxy emission and 
the effects of variable seeing for extended sources. 
Right panel: The structure functions of the $5100\ \rm{\AA}$ continuum 
for the SDSS-RM point-source sample, with galaxy emission subtracted 
using the spectral decomposition of \cite{shen15b} (and removing 
sources with $>50\%$ host contribution). After the host-galaxy effects 
are removed, sources in low-luminosity bins show higher variability.}
\label{fig:sf5k}
\end{figure*}

We discuss the structure function of the 
$5100\ \rm{\AA}$ continuum for SDSS-RM quasars. Since 
now we are dealing with $z<0.8$ quasars, we slightly 
changed the quasar luminosity bins: $L_{\mathrm{Bol}}< 
10^{44.8}\ \mathrm{erg\ s^{-1}}$; $10^{44.8}\ \mathrm{erg\ 
s^{-1}} \leq L_{\mathrm{Bol}}< 10^{45.2}\ \mathrm{erg\ 
s^{-1}}$; $10^{45.2}\ \mathrm{erg\ s^{-1}} \leq 
L_{\mathrm{Bol}}<10^{45.6}\ \mathrm{erg\ s^{-1}}$; 
$L_{\mathrm{Bol}} > 10^{45.6}\ \mathrm{erg\ s^{-1}}$. 
We calculated the structure function for each bin. Our 
results are shown in the left panel of Figure~\ref{fig:sf5k}. 
As expected, the structure functions increase with $\Delta t$. 
In contrast to the structure functions of the $3000\ \rm{\AA}$ 
continuum, the structure functions of the $5100\ \rm{\AA}$ 
continuum do not appear to be a monotonic function of 
quasar luminosity. 

Interpreting the structure function of the $5100\ \rm{\AA}$ 
continuum is challenging because the host-galaxy contribution 
cannot be neglected. The host-galaxy emission is not 
expected to vary on these timescales. Therefore, a significant 
host contribution would act to dilute the measured $\Delta m$. 
On the other hand, if the host-galaxy emission is extended 
with respect to the spectroscopic fiber (which has a diameter 
of $2^{''}$), seeing variations will cause apparent 
variability of the host-galaxy stellar light and increase  
the observed $\Delta m$ (see Appendix~\ref{sec:o3}). 

We first restrict the sample to point sources, as classified by 
the SDSS imaging, to eliminate luminous host galaxies. 
Restricting to point sources avoids the additional variability 
due to variable seeing. We then restricted the sample to quasars 
with spectral decomposition performed by \cite{shen15b} and 
subtracting the estimated host contribution to the $5100\ \rm{\AA}$ 
continuum (quasars with $>50$\% galaxy light are rejected). 
Our results are shown in the right panel of Figure~\ref{fig:sf5k}. 

Comparing the host-corrected and uncorrected $5100\ \rm{\AA}$ 
continuum structure functions, it is evident that the 
galaxy emission does have significant effects: ``host-subtracted'' 
structure functions increase due to the removal of host dilution. 
Also, after applying the correction, the structure functions tend to 
decrease with $L_{\rm{Bol}}$. Some previous works 
\citep[e.g.,][]{kelly09} did not account for host-galaxy emission, 
and therefore may have underestimated quasar variability at 
longer wavelengths. However, the effects of host-galaxy contamination 
are likely to be much smaller for very luminous SDSS quasars 
(e.g., M10). 

\section{Structure Functions of Emission lines}
\label{sec:sfeml}
In Section~\ref{sec:sfcont}, we found that the variability 
of the continuum emission (at $3000\ \rm{\AA}$) is 
consistent with the DRW model. In this section, we 
present the structure functions of broad \MgII \ and \Hbeta . 
The study of the variability of broad emission lines provides 
information about the transfer function governing the 
response of the broad line region to incident continuum 
emission. The transfer function, in turn, reflects the 
structure and ionization state of the BLR.

\subsection{\rm{\MgII}}
\label{sec:sfmg}
\begin{figure*}
\epsscale{1.2}
\plotone{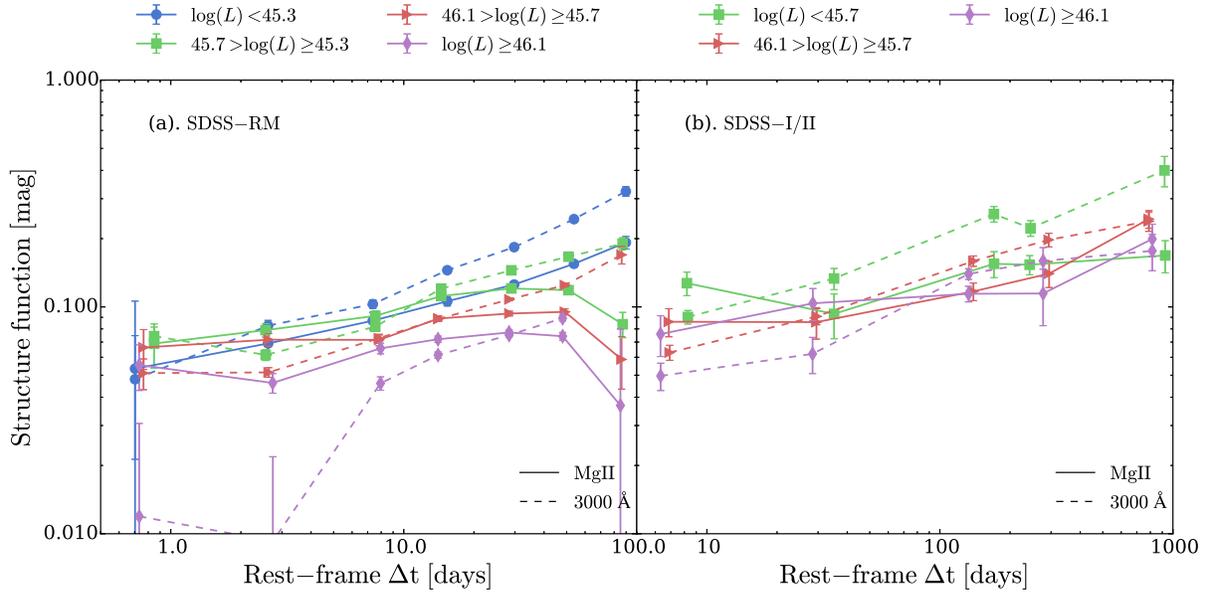}
\caption{Left panel: The structure functions of \MgII \ (solid 
lines) for SDSS-RM sources. For comparison, we also present 
the structure functions of the $3000\ \rm{\AA}$ continuum (only 
using sources with well-measured \MgII ). It is evident that the 
structure functions of \MgII \ depend on $L_{\rm Bol}$ in the 
same way as the $3000\ \rm{\AA}$ continuum. This result 
is consistent with the idea that the variability of \MgII \ is 
driven by continuum variability. However, the $3000\ \rm{\AA}$ 
continuum is more variable than \MgII . Also, the slopes of the 
\MgII \ and $3000\ \rm{\AA}$ continuum structure functions are 
different (the structure function of \MgII \ is shallower in 
each luminosity bin). Note that the apparent variability decrease 
at $\Delta t\sim 100$ days is likely due to the fact that the 
spectrophotometric errors dominate over the intrinsic variability. 
Right panel: The structure functions of \MgII \ (solid lines) 
for SDSS-I/II sources. For comparison, we also present the 
structure functions of the $3000\ \rm{\AA}$ continuum (only 
using sources with well-measured \MgII ). Considering the fact 
that SDSS-I/II sources are generally more luminous than SDSS-RM 
sources, we only created three luminosity bins. }
\label{fig:sfmgii}
\end{figure*}

We first consider the variability of \MgII \ for SDSS-RM sources. 
We again divided our sources into the same four $L_{\rm{Bol}}$ 
bins as we did for the $3000\ \rm{\AA}$ continuum and calculated 
the structure function for each bin. We binned quasars by the 
continuum luminosity rather than the line luminosity so that 
it is easier to compare to the structure function of continuum 
emission. Our results are presented in the left panel of 
Figure~\ref{fig:sfmgii}. 

Compared to the $3000\ \rm{\AA}$ continuum, the structure 
functions of \MgII \ also increase with $\Delta t$ and (weakly) 
decrease with $L_{\rm{Bol}}$. We argue that this similarity is 
consistent with the idea that the variability of \MgII \ is driven 
by the variability of the continuum emission \citep[see the cross 
correlation analysis of][]{shen15c}. 

The structure functions of \MgII \ also differ from those of 
the $3000\ \rm{\AA}$ continuum in several aspects. First, 
the variability amplitude of the $3000\ \rm{\AA}$ continuum 
is generally larger than that of \MgII , 
for each luminosity bin. In addition, the difference between 
the variability amplitudes of the $3000\ \rm{\AA}$ continuum 
and those of \MgII \ decrease with $L_{\rm Bol}$ (similar 
to the results of Section~\ref{sec:lcv}). We also found the 
shapes of the structure functions of \MgII \ and those of the 
$3000\ \rm{\AA}$ continuum are different by performing the 
same statistical hypothesis test we did in Section~\ref{sec:sf3k}. 
We found that, for all four luminosity bins, the null 
hypothesis is rejected (i.e., the $p$ value of the chi-squared 
test is $\ll 0.01$). That is, our data indicate different 
slopes for the \MgII \ and the $3000\ \rm{\AA}$ continuum 
structure functions. Note that \MgII \ typically has larger 
measurement errors, and we expect it suffers from a stronger 
bias than the $3000\ \rm{\AA}$ continuum. To ensure the 
observed differences in shapes are not caused by the additional 
bias, we performed the ``DRW simulation'' described in 
Section~\ref{sec:sf3k} for \MgII . We found that this simulation 
cannot reproduce the observed structure functions of \MgII \ 
(both the normalization factors and shapes). Therefore, we 
conclude that the differences between the slopes for the \MgII \ 
and the $3000\ \rm{\AA}$ continuum structure functions are 
intrinsic. 

The right panel of Figure~\ref{fig:sfmgii} shows the \MgII \ 
structure functions for the SDSS-I/II quasars. The variability 
amplitude of \MgII \ is again smaller than that of the $3000\ 
\rm{\AA}$ continuum for each luminosity bin. We also 
compared the shapes of the structure functions of \MgII \ 
and those of the $3000\ \rm{\AA}$ continuum by performing 
the same hypothesis test. We found that, for the two 
low-luminosity bins, the shapes of the structure functions of 
\MgII \ and those of the $3000\ \rm{\AA}$ continuum are 
different (i.e., the $p$ value is $\ll 0.01$). For the highest 
luminosity bin (i.e., the $\log L_{\rm{Bol}}>46.1$ bin), the 
$p$ value under the null hypothesis is $0.012$, and we 
cannot reject the idea that the shapes are the same. If 
we restrict the comparison to rest-frame timescales 
$\Delta t>100\ \rm{days}$ (the timescales that are not 
covered by the SDSS-RM data), we found that we cannot 
reject the null hypothesis and the \MgII \ and the $3000\ 
\rm{\AA}$ continuum structure functions have consistent 
shapes in all luminosity bins (the $p$ value under the null 
hypothesis is $\gg 0.01$). The statistical test at $\Delta 
t>100$ days is limited, however, by having only three data 
points in each luminosity bin and therefore lacks power to 
reject the null hypothesis (the difference in slope $\beta$ 
is poorly constrained to be $\lesssim 0.5$).

\subsection{\Hbeta}
\label{sec:sfhb}
\begin{figure}
\epsscale{1.2}
\plotone{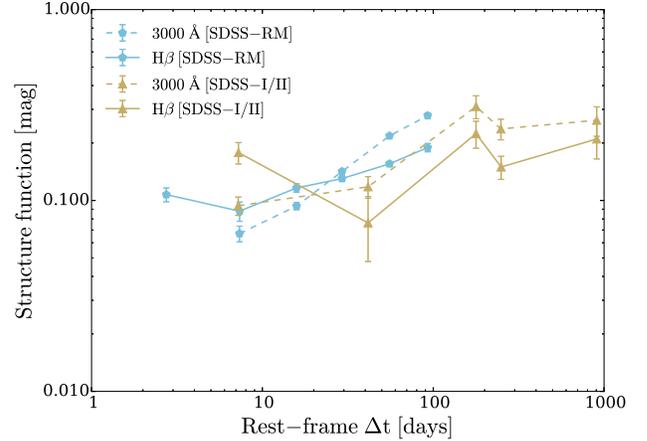}
\caption{The structure functions of \Hbeta \ for SDSS-RM sources 
(pentagons) and SDSS-I/II sources (triangles). As a comparison, 
we also included the structure functions of the $3000\ \rm{\AA}$ 
continuum (dashed lines). \Hbeta \ is less variable than the $3000\ 
\rm{\AA}$ continuum. The shapes of the structure functions of 
\Hbeta \ and the $3000\ \rm{\AA}$ continuum are different on 
timescales $\Delta t< 100\ \rm{days}$. On longer timescales, 
the shapes are similar. Note that only quasars with 
``well-measured''  \Hbeta \ and $3000\ \rm{\AA}$ continuum 
are included.}
\label{fig:sfhb}
\end{figure}

In this section, we study the structure functions of \Hbeta \ 
for the SDSS-RM and SDSS-I/II data. In order to compare the 
structure functions of \Hbeta \ to that of the $3000\ \rm{\AA}$ 
continuum, we only consider sources with ``well-measured'' (i.e., 
satisfying our selection criteria) \Hbeta \ and $3000\ \rm{\AA}$ 
continuum. We choose the $3000\ \rm{\AA}$ continuum instead of 
the $5100\ \rm{\AA}$ continuum because the latter is significantly 
affected by galaxy emission (see Section~\ref{sec:sf5k}). Due to 
the limited total number of sources, we do not divide our sources 
by luminosity. Figure~\ref{fig:sfhb} shows our results. 

It is evident that, for both SDSS-RM and SDSS-I/II sources, 
the variability of \Hbeta \ is smaller than that of the $3000\ 
\rm{\AA}$ continuum on timescales of rest-frame $\Delta t>20\ 
\rm{days}$. We compared the shape of the structure function 
of \Hbeta \ with that of the $3000\ \rm{\AA}$ continuum by 
performing the same hypothesis test we did for \MgII . From 
this test, the shapes of the \Hbeta \ and continuum structure 
functions are significantly different (i.e., the $p$ value 
under the null hypothesis is $\ll 0.01$). However, if we 
again restrict the comparison to rest-frame $\Delta t>100\ 
\rm{days}$, the hypothesis that the shapes of the structure 
functions of \Hbeta \ and the $3000\ \rm{\AA}$ continuum 
are the same cannot be rejected (the $p$ value under the 
null hypothesis is $\gg 0.01$). Again, the statistical test 
at $\Delta t>100$ days is limited (the ``allowed'' difference 
in slope $\beta$ is only constrained to be $\lesssim 0.8$). 

\begin{figure}
\epsscale{1.2}
\plotone{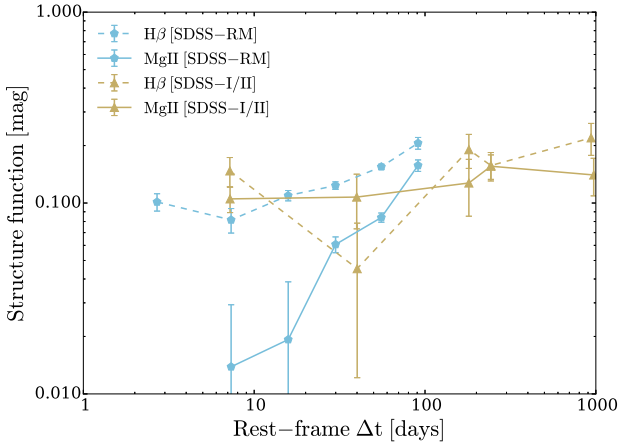}
\caption{A comparison of the structure functions of \MgII \ 
with those of \Hbeta \ for both SDSS-RM (pentagons) and SDSS-I/II 
(triangles) sources. For the SDSS-RM sources, the shapes of 
the structure functions of \MgII \ and those of \Hbeta \ are 
different. For the SDSS-I/II sources, our statistical hypothesis 
test indicates that the shapes of the structure functions of 
\MgII \ and those of \Hbeta \ are not significantly different. 
Note that only quasars with ``well-measured'' \MgII \ and \Hbeta \ 
are included.}
\label{fig:sfmg-hb}
\end{figure}

We also compare the structure function of \MgII \ with that 
of \Hbeta . Only $26$ sources with ``well-measured'' 
\MgII \ and \Hbeta \ are included. Due to the limited sample 
size, we do not bin by luminosity. Our results 
are presented in Figure~\ref{fig:sfmg-hb}. For the SDSS-RM 
sample, the variability amplitude of \Hbeta \ is larger than 
that of \MgII , which was also indicated by the analysis 
described in Section~\ref{sec:lcvar} 
(Figure~\ref{fig:hbeta-mgii-lcvar}). For the SDSS-I/II sample, 
the difference is not significant. 

We again performed the same hypothesis test to test whether 
the structure function of \MgII \ and that of \Hbeta \ share 
the same shape. For the SDSS-RM sample, our test indicates 
that the \MgII \ and \Hbeta \ structure functions have 
statistically different shapes (i.e., the $p$ value under 
the null hypothesis is $< 0.01$). For the SDSS-I/II sample, 
the null hypothesis cannot be rejected by our data (i.e., 
the $p$ value under the null hypothesis is $\gg 0.01$, and 
the difference in slope $\beta$ is $\lesssim 0.3$). 

\section{Discussion: The Nature of Quasar Variability}
\label{sec:disall}
\subsection{Quasar variability and quasar properties}
\label{sec:dis1}
As revealed by many previous investigations \citep[e.g.,][]{kelly09, 
mac12}, quasar variability is controlled by quasar properties 
(see Eq.~\ref{eq:m10}). In Figure~\ref{fig:sf3k}, we showed that 
the M10 relations can effectively reproduce our results. Therefore, 
our results are consistent with the idea that quasar properties 
(e.g., $L_{\rm Bol}$ and $M_{\rm BH}$) determine the structure 
of the accretion disk which in turn controls the instabilities in 
the accretion disk. 

The thermal timescale of the accretion disk is \citep[e.g.,][]{kat98, 
kelly09} 
\begin{equation}
t_{\rm{TH}}\approx 53.1\times(\frac{\alpha}{0.01})^{-1} 
(\frac{M_{\rm{BH}}}{10^8\ M_{\odot}}) 
(\frac{R}{10\ R_{\rm{S}}})^{3/2}\ \rm{days} \\, 
\end{equation}
where $\alpha$ is the dimensionless viscosity parameter \citep{sha73}, 
and $R_{\rm{S}}=2GM_{\mathrm{BH}}/c^2$ is the Schwarzschild radius. 
Assuming the $3000\ \rm{\AA}$ continuum is produced at $\sim 30\ R_{\rm{S}}$ 
\citep[based on microlensing observations, see Eq.~4 of][]{mor10}, 
the expected thermal timescale is around $174\ \rm{days}$ (for 
$\alpha=0.05$ and $M_{\mathrm{BH}}=10^{8.5}\ M_{\odot}$ which is the 
median value for our sample). The estimated $\tau$ ($\sim 
199.5^{+116.7}_{-73.6}$ days, see Section~\ref{sec:sf3k}) from the 
SDSS-I/II data is consistent with the thermal timescale of the 
accretion disk. Therefore, the DRW model can be explained by thermal 
fluctuations in the accretion disk \citep[see also][]{kelly09}. On 
timescales much smaller than $\tau$, the thermal fluctuations 
\citep[which are ultimately driven by magnetic turbulence, see 
recent numerical simulations, e.g.,][and also a detailed theoretical 
calculation by Lin et al. 2012]{hir09, jiang13} in the accretion 
disk drive the random-walk-like fluctuations in $L$. On longer 
timescales ($\gg \tau$), the disk can adjust itself to the thermal 
fluctuations and therefore the fluctuations in $L$ are purely white 
noise and/or are related to fluctuations in $\dot{M}$ \citep[which 
correspond to the much longer viscous timescale, see e.g.,][]{lyu97, 
kelly11}. 

It has been found that, on short timescales ($\sim$ days), the 
variability of quasar continuum emission is smaller than the 
DRW model predicts \citep[e.g.,][]{mus11, zu13, kas15}. Although 
SDSS-RM includes some variability data on these short timescales, 
the median spectroscopic sampling interval over the campaign was 
$\sim 4$ days in observed frame (i.e., $\gtrsim 1.3$ days in rest 
frame), and the large measurement and spectrophotometric errors 
make it difficult to place meaningful constraints on variability 
over $\Delta t \lesssim 5$ days. In particular, on very short 
timescales (where the intrinsic variability is small), the observed 
variability is dominated by the measurement and spectrophotometric 
errors rather than the intrinsic variability, and the structure 
function estimation suffers from significant bias (as illustrated 
in the middle panel of Figure~\ref{fig:sf3k}).

\subsection{Implications for Reverberation-Mapping 
Projects}
\label{sec:dis2}
In the SDSS-RM overview paper, \cite{shen15} simulated the 
expected results from the SDSS-RM project. In this section, 
we explore the validity of these simulations and the 
implications of our results for future reverberation-mapping 
campaigns. 

We start with the model used to generate quasar continuum 
light curves. In \cite{shen15}, they generated continuum light 
curves based on the DRW model and the M10 relations. As shown 
in Section~\ref{sec:sf3k}, the DRW model indeed describes the 
rest-frame $10-100\ \rm{day}$ variability of quasar continuum 
emission well. The M10 relations can also effectively reproduce 
the structure function, although low-luminosity quasars may 
actually be slightly more variable (by a factor of $\sim 1.3$) 
than predicted by extrapolating the M10 relations. Therefore, 
the initial SDSS-RM simulations of quasar continuum light curves 
are likely to be valid. 

Justifying the transfer function is difficult since we actually 
cannot constrain its exact shape. However, assuming the width 
of the transfer function to be $10\%$ of the time lag (i.e., not 
a $\delta$ function) is consistent with our results. We find 
that the transfer function is likely to have a width in the range 
of $1-100$ days (see Section~\ref{sec:dis3}).  Our loose constraints 
on the transfer-function width are consistent with width estimates 
from velocity-resolved reverberation mapping results 
\citep[e.g.,][]{gri13, de15}. 

Last but not least, we find that, for a large population of quasars, 
\MgII \ varies significantly. Like \Hbeta , \MgII \ is likely to 
respond to the continuum emission (as they depend on quasar luminosity 
in a similar way) and therefore has the potential to be used for 
reverberation-mapping campaigns.

\subsection{The Broad Emission Line Transfer 
Function}
\label{sec:dis3}
The detailed ionization state, dynamical structure, and kinematic 
motions of the BLR are still poorly constrained from observations. 
In Sections~\ref{sec:lcv} and \ref{sec:sfeml}, we compared the 
variability of two broad emission lines (\MgII \ and \Hbeta ) 
with that of $L_{3000}$. We demonstrated that both broad lines 
were less variable than the continuum emission (at fixed time 
separation and luminosity).  In addition, the structure functions 
of the broad emission lines are shallower than the $3000\ \rm{\AA}$ 
continuum structure functions, indicating that the broad emission 
lines are not driven by the same DRW model which describes the 
variability of the $3000\ \rm{\AA}$ continuum.  \Hbeta \ is 
typically more variable than \MgII . In this section, we discuss 
the implications of these results for our understanding 
of BLR. 

Let us first define a transfer function $\Phi (t_s)$ which 
governs the response of the emission-line light curve $f_l(t)$ 
to the incident continuum emission $f_c(t)$, after a light-travel 
time delay $t_s$ \citep[e.g.,][]{bla82}:
\begin{equation}
\label{eq:transf}
f_l(t)=\int dt_s \Phi (t_s)f_c(t-t_s)
\end{equation}
where $f_c(t)$ is the light curve of the continuum emission. 
The broad-line structure 
functions are flatter than those of the continuum (i.e., $3000\ 
\rm{\AA}$, whose variability is consistent with the DRW model). 
If the variability of the EUV continuum is similar to the variability 
of the $3000\ \rm{\AA}$ continuum, our results demonstrate that 
the transfer functions of the broad emission lines are not consistent 
with the $\delta$-function: a broad (in time) transfer function 
would effectively flatten the input structure function. \MgII \ 
and \Hbeta \ are driven by ionizing photons of $E>15\ \rm{eV}$ 
and $E>13.6\ \rm{eV}$, respectively.  This higher-energy flux 
likely has a shorter $\tau$ than than the $3000\ \rm{\AA}$ continuum, 
leading to an apparently flatter structure function. However, 
extrapolating the M10 relations to $E\sim 15\ \rm{eV}$, the expected 
$\tau$ is still $\sim 100\ \rm{days}$.  

The shape differences disappear when we only consider variability 
on timescales of rest-frame $\Delta t>100\ \rm{days}$. This 
conclusion, if it is robust, suggests that the width of the 
transfer function for \MgII \ is (for our SDSS-RM sample) 
less than $100\ \rm{days}$. The same conclusion holds for 
\Hbeta . Therefore, on longer timescales, the variability of 
\MgII \ (or \Hbeta ) has similar timescale-dependence to the 
variability of the $3000\ \rm{\AA}$ continuum. 

Our data demonstrate that both \MgII \ and \Hbeta \ have lower 
variability amplitude than the $3000\ \rm{\AA}$ continuum. This 
result is also consistent with \cite{mac12} (see their Figure~$13$) 
who (indirectly) find that \MgII \ is less variable than the local 
continuum by exploring the photometric variability as a function 
of rest-frame wavelength \citep[see also][]{ive04}. The 
EUV continuum which actually drives both lines is probably more 
variable than the $3000\ \rm{\AA}$ continuum, since variability 
increases with decreasing wavelength \citep[e.g.,][]{vb04,mac10}. 
Indeed, early EUV observations reveal strong variability (a 
factor of $2$) even within $1\ \rm{day}$ \citep[e.g.,][]{mar97, 
utt00, hal03}. Thus the amplitude of the transfer function is 
likely to be significantly less than one, with the broad lines 
less variable than their incident continuum. This result is 
consistent with detailed photoionization calculations 
\citep[e.g.,][]{kor00, kor04}. 

Our data suggest that the difference between the $3000\ \rm{\AA}$ 
continuum and \MgII \ variability decreases with $L_{\rm Bol}$. 
That is, as quasar luminosity decreases, the variability amplitude 
of \MgII \ increases more slowly than $3000\ \rm{\AA}$. This may 
be related to a changing ionization structure as luminosity changes: 
this effectively changes the radius of the BLR (as inferred by 
reverberation-mapping studies, which measure a responsivity-weighted 
radius), and the responsivity (defined as the ratio of the variability 
of emission lines to that of continuum emission) increases with 
increasing radius \citep{kor04}. 

We find that \Hbeta \ is slightly more variable than \MgII , in agreement 
with previous quasar spectral variability studies \citep{kok14}. The 
ionizing continuum that drives \Hbeta \ ($E>13.6\ \rm{eV}$) is similar 
to that driving \MgII \ ($E>15\ \rm{eV}$). Therefore, the differences 
between \MgII \ and \Hbeta \ variability are not likely to be due to 
the differences in ionizing continuum. 

The variability differences might be caused by the fact that \Hbeta 
\ is a recombination line while \MgII \ is a collisionally excited 
resonance line \citep[e.g.,][]{mac72, net80}. Each of these processes 
depends on the changes in incident EUV flux in a different way. This 
difference might lead to a lower variability of \MgII \ than \Hbeta . 

The differences between \MgII \ and \Hbeta \ might also be caused 
by the different optical depth for the two emission lines. Since 
\MgII \ is a resonance line, its optical depth is expected to be 
larger than that of \Hbeta . At a sufficiently high density, 
\MgII \ photons will be absorbed and re-emitted many times before 
escaping the BLR. This process might effectively stabilize 
changes in \MgII \ luminosity by diluting the response to continuum 
changes. 

The difference between \MgII \ and \Hbeta \ could also indicate 
the structure of the BLR that emits \MgII \ is different from 
the BLR that produces \Hbeta . For instance, it is possible 
the BLR material that emits \MgII \ extends to larger radius 
that that of \Hbeta . Therefore, as the quasar continuum varies, 
\MgII \ responds on a wider range of timescales, and the variability 
of \MgII \ is again diluted. 

Some of these ideas are included in the ``local optimally-emitting 
cloud'' (LOC) model which assumes that the BLR consists of various 
density gas clouds. A detailed photoionization equilibrium calculation 
of the LOC model reveals that \MgII \ is less variable than 
\Hbeta \ over most of the typical range of BLR conditions 
\citep{kor97, kor00, kor04}. 

Our data also indicate that the shapes of the structure 
functions of \MgII \ and \Hbeta \ are different at $\Delta t<100$ 
days. These results suggest that the transfer 
functions of \MgII \ and \Hbeta \ are also different in 
width: that is, the transfer functions of \MgII \ and \Hbeta \ 
have different radial profiles. At longer timescales ($\Delta 
t>100$ days), the shapes of the structure functions of \MgII \ 
and \Hbeta \ are similar (the difference in slope $\beta$ is 
$\lesssim 0.3$). This result is expected since the widths 
of the transfer functions for \MgII \ and \Hbeta \ are both 
likely less than $100$ days. 

Our comparison of \Hbeta \ and \MgII \ variability is largely 
limited to low-luminosity quasars due to the low-redshift 
requirement for each spectrum to include both lines. It is 
possible that the relative difference in responsivity 
between \Hbeta \ and \MgII \ has a luminosity dependence 
\citep{kor97, kor00,kor04}, and comparing the different 
variability behavior of \Hbeta \ and \MgII \ over a larger 
luminosity range would place additional constraints on the 
physical conditions and excitation sources of the BLR.

\section{Summary}
\label{sec:summary}
Using SDSS-RM and SDSS I/II data, we studied the variability 
of continuum emission probed at $3000\ \rm{\AA}$ and $5100\ 
\rm{\AA}$ continuum and the variability of the \MgII \ and 
\Hbeta \ broad emission lines. 
Our results can be summarized as follows: 
\begin{enumerate}
\item We determined the variability of \MgII \ for a large quasar 
sample. We found that, like \Hbeta , \MgII \ varies, consistent 
with the scenario that \MgII \ varies in response to the variability 
of the continuum emission (Figures~\ref{fig:mgii-lcvar} and 
\ref{fig:sfmgii}; Sections~\ref{sec:lcv} and~\ref{sec:sfeml}). 

\item We found that the shapes of the structure functions of the 
$3000\ \rm{\AA}$ continuum and those of broad emission lines are 
different, indicating the transfer functions governing the response 
of broad emission lines are broad in time. We also found that the 
difference between the variability of the $3000\ \rm{\AA}$ continuum 
and broad emission lines decreases with quasar luminosity 
(Figures~\ref{fig:mgii-l3k-lcvar} and \ref{fig:sfmgii}; 
Sections~\ref{sec:lcv} and~\ref{sec:sfeml}), consistent 
with photoionization model predictions of \cite{kor04}. 

\item We confirmed that the continuum variability on timescales of 
$\Delta t>5$ days is well-described by the DRW model (the difference 
in slope $\beta\lesssim 0.1$, see Figure~\ref{fig:sf3k} and 
Section~\ref{sec:sf3k}), and that the continuum variability is a 
function of quasar properties (Figure~\ref{fig:sf3k}. 
Section~\ref{sec:dis1}).

\item Emission from the host galaxy introduces a significant bias 
to the measured quasar variability at rest-frame $5100\ \rm{\AA}$ 
(Figure~\ref{fig:sf5k}; Section~\ref{sec:sf5k}). 

\item We also found that the structure functions of \MgII \ and 
\Hbeta \ have statistically different shapes (Figure~\ref{fig:sfmg-hb}; 
Sections~\ref{sec:sfhb} and~\ref{sec:dis3}). Also, \Hbeta \ is 
slightly more variable than \MgII , consistent with the predictions 
of the LOC model \citep[e.g.,][]{kor00,kor04}. Our results may 
be explained by the fact that the two broad emission lines have 
different radiative mechanisms, geometrical configurations, and/or 
optical depths (see Sections~\ref{sec:lcv} and~\ref{sec:dis3}). 
\end{enumerate}

Our results indicate that the predictions of the SDSS-RM project 
are accurate (see Section~\ref{sec:dis2}). Continuing observations 
should enable accurate estimation of $M_{\rm BH}$ for a large set 
of $z>1$ quasars utilizing the reverberation-mapping technique 
\citep[using both \Hbeta \ and \MgII , see][]{shen15c}.

\acknowledgments
We thank Christopher Kochanek, Michael Eracleous, Michael Goad, 
Kirk Korista, and Guang Yang for beneficial discussions. We thank 
the referees for their constructive comments. MYS 
acknowledges support from the China Scholarship Council 
(No.~[2013]3009). JRT and YS acknowledge support from NASA 
through Hubble Fellowship grants \#51330 and \#51314, respectively, 
awarded by the Space Telescope Science Institute, which is 
operated by the Association of Universities for Research in 
Astronomy, Inc., for NASA under contract NAS 5-26555. 
WNB acknowledges support from NSF grant AST-1108604 and the 
V. M. Willaman Endowment. KDD is supported by an NSF AAPF 
fellowship awarded under NSF grant AST-1302093. 

Funding for SDSS-III has been provided by the Alfred P. Sloan 
Foundation, the Participating Institutions, the National Science 
Foundation, and the U.S. Department of Energy Office of Science. 
The SDSS-III web site is \url{http://www.sdss3.org/}. Funding for 
the SDSS and SDSS-II has been provided by the Alfred P. Sloan 
Foundation, the Participating Institutions, the National Science 
Foundation, the U.S. Department of Energy, the National Aeronautics 
and Space Administration, the Japanese Monbukagakusho, the Max 
Planck Society, and the Higher Education Funding Council for 
England.

SDSS-III is managed by the Astrophysical Research Consortium for the 
Participating Institutions of the SDSS-III Collaboration including the 
University of Arizona, the Brazilian Participation Group, Brookhaven 
National Laboratory, Carnegie Mellon University, University of Florida, 
the French Participation Group, the German Participation Group, Harvard 
University, the Instituto de Astrofisica de Canarias, the Michigan 
State/Notre Dame/JINA Participation Group, Johns Hopkins University, 
Lawrence Berkeley National Laboratory, Max Planck Institute for 
Astrophysics, Max Planck Institute for Extraterrestrial Physics, New Mexico 
State University, New York University, Ohio State University, Pennsylvania 
State University, University of Portsmouth, Princeton University, the 
Spanish Participation Group, University of Tokyo, University of Utah, 
Vanderbilt University, University of Virginia, University of Washington, 
and Yale University.

\appendix
\section{STRUCTURE FUNCTION ESTIMATORS}
\label{sec:sfest}
The structure function measures the statistical dispersion of 
two observations separated by $\Delta t$. Several popular 
non-parametric statistical dispersion estimators have been 
proposed: the interquartile range (IQR), the median absolute 
deviation (MAD), the average absolute deviation (AAD), and 
the standard deviation. Their definitions are:
\begin{itemize}
\item the IQR estimator \citep[used by][]{mac12}: see 
Section~\ref{sec:def};

\item the MAD estimator: 
\begin{equation}
\mathrm{SF_{MAD}}(\Delta t) = \sqrt{(1.48\mathrm{MAD}(\Delta m))^2 
-\widetilde{\sigma^2_{\rm{e}}}} \nonumber \\,
\end{equation}
where $\mathrm{MAD}(\Delta m)=\widetilde{\left|\Delta m - 
\widetilde{\Delta m}\right|}$;

\item the AAD estimator \citep[used by][]{vb04}:
\begin{equation}
\mathrm{SF_{AAD}}(\Delta t)= \sqrt{\frac{\pi}{2}\left<\left|\Delta m 
\right|\right>^2 -\left<\sigma^2_{\rm{e}}\right>} \nonumber \\.
\end{equation}

\item the maximum-likelihood standard deviation (ML) estimator: 
the standard deviation can be measured using the maximum-likelihood 
method proposed by \cite{alm00} (see their Section~3.1). 
\end{itemize}

We subtract the spectrophotometric error from each structure 
function calculated from the stars using the same estimator. 
The constant factors in each of the four structure-function 
estimators normalizes them to be equivalent to the standard 
deviation of a Gaussian distribution (assuming Gaussian 
measurement and spectrophotometric errors). Our data and 
uncertainties have non-Gaussian tails, which lead to differences 
in the estimators: for distributions with larger tails than 
a Gaussian, the IQR and MAD estimators are smaller than the 
AAD and standard deviation estimators.  

\begin{figure}
\epsscale{1.2}
\plotone{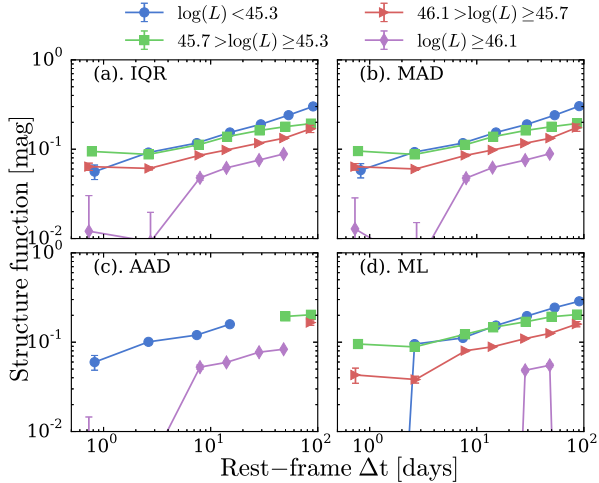}
\caption{The structure function of the $3000\ \rm{\AA}$ continuum 
for each quasar luminosity bin. The IQR, MAD, AAD, and ML estimators 
are shown. The IQR and MAD estimators are stable against 
measurements with relatively large uncertainties and also robust 
against outliers. 
}
\label{fig:sf3k-full}
\end{figure}

The four structure functions of the $3000\ \rm{\AA}$ continuum 
are plotted in Figure~\ref{fig:sf3k-full}. Note that if the 
number of flux pairs for a given $\Delta t$ bin is less than 
$10$, we do not calculate the variability in this particular 
bin. For each $L_{\rm{Bol}}$ bin with many quasars, each quasar 
has different intrinsic variability and measurement/spectrophotometric 
errors. That is, the ensemble distribution of $\Delta m$ is 
a superposition of many Gaussian distributions and is not a 
perfect Gaussian but instead has exponential tails. Moreover, 
the subtraction of measurement and spectrophotometric errors 
(i.e., subtract $\tilde{\sigma_e}$ or $\left< \sigma_e 
\right>$) has an effect on the estimation of the ensemble 
structure function. This results in slightly different results 
for each of the four \mbox{structure-function} estimators. 

For the AAD estimator, since we are subtracting the mean 
value of measurement and spectrophotometric errors, our 
results may be biased by quasars with measurement and 
spectrophotometric errors that are much larger than the 
intrinsic variability. The standard deviation estimator uses 
individual errors (using the maximum likelihood method), 
and so is stable against quasars with relatively large 
measurement and spectrophotometric errors. The IQR and 
MAD estimators use median values of the measurement 
and spectrophotometric errors, and so are also stable against 
quasars with large measurement and spectrophotometric 
errors. Moreover, the IQR and MAD estimators are also robust 
against outliers in the distribution of $\Delta m$. 
 
The IQR, MAD and ML (or standard deviation) estimators give 
similar results: (1) the ensemble structure function increases 
with $\Delta t$ in a similar way; (2) the ensemble structure 
function decreases with quasar luminosity in a similar way. 

\begin{figure}
\epsscale{1.2}
\plotone{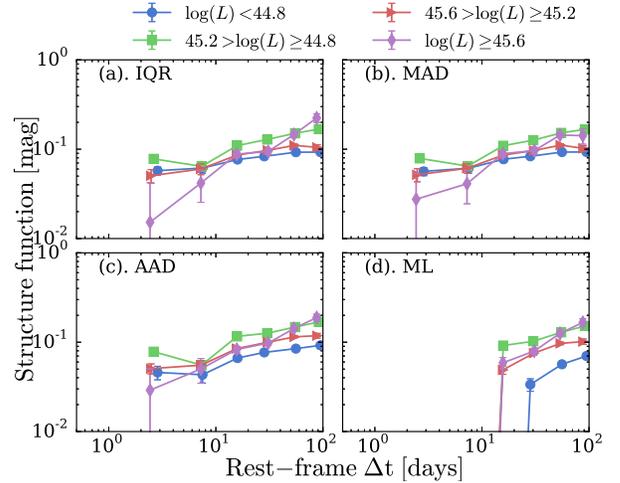}
\caption{The structure function of the $5100\ \rm{\AA}$ continuum 
for each luminosity bin. The IQR, MAD, AAD, and ML estimators 
are shown.}
\label{fig:sf5k-full}
\end{figure}

\begin{figure}
\epsscale{1.2}
\plotone{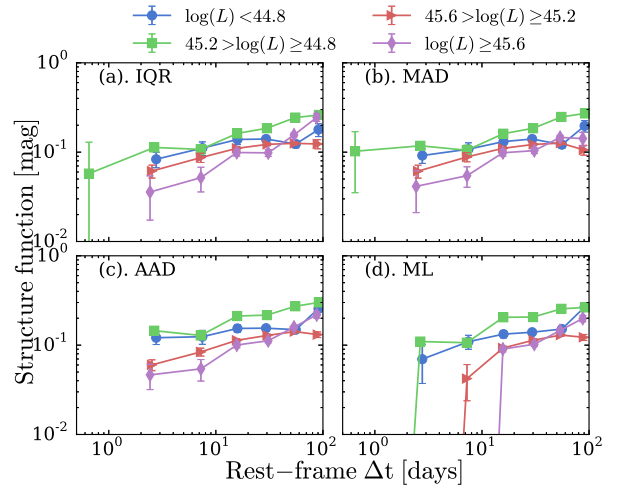}
\caption{The structure function of the $5100\ \rm{\AA}$ continuum 
for each luminosity bin. The IQR, MAD, AAD, and ML estimators 
are shown. Only point sources are included, and galaxy 
emission is subtracted.}
\label{fig:sf5k-full2}
\end{figure}

The four structure functions of the $5100\ \rm{\AA}$ continuum are 
plotted in Figure~\ref{fig:sf5k-full} for the full ``clean'' sample of 
SDSS-RM quasars. Figure~\ref{fig:sf5k-full2} shows only point 
sources with galaxy emission subtracted, using the spectral 
decomposition of \cite{shen15b} and removing quasars with $> 
50\%$ host contribution at $5100\ \rm{\AA}$.  

\begin{figure}
\epsscale{1.2}
\plotone{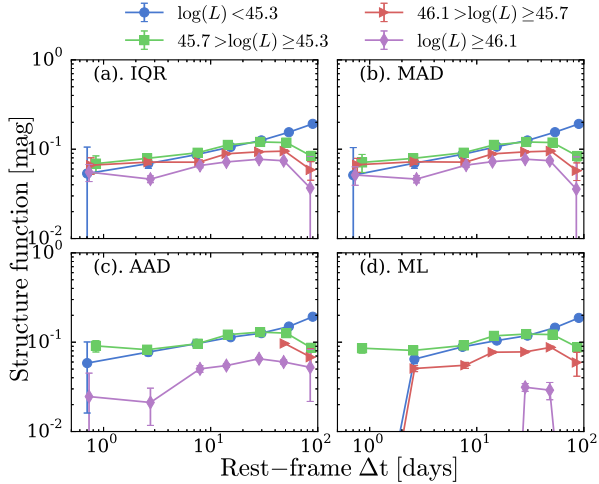}
\caption{The structure function of \MgII \ for each luminosity 
bin. The IQR, MAD, AAD, and ML estimators are shown.}
\label{fig:sfmg-full}
\end{figure}

\begin{figure}
\epsscale{1.2}
\plotone{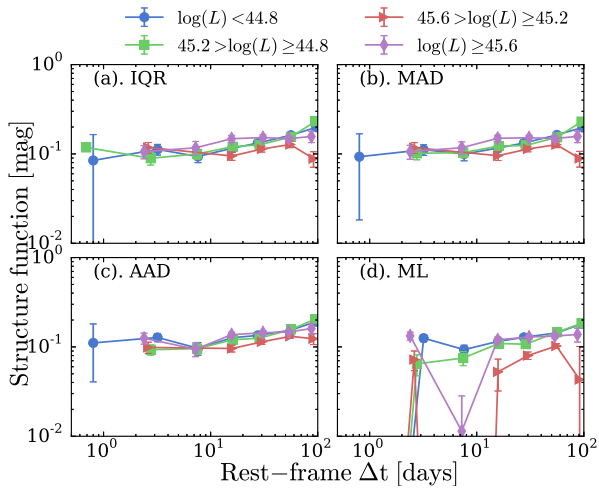}
\caption{The structure function of \Hbeta \ for each luminosity 
bin. The IQR, MAD, AAD, and ML estimators are shown.}
\label{fig:sfhb-full}
\end{figure}

The four structure functions of \MgII \ and \Hbeta \ are shown in 
Figures~\ref{fig:sfmg-full} and \ref{fig:sfhb-full}, respectively. In 
all four cases, the broad-line structure functions are flatter than 
the continuum structure functions. 

As shown in this section, the IQR and MAD estimators are robust 
against outliers. We prefer the IQR since it has been adopted in 
\cite{mac12}. 

\section{SDSS-RM SPECTROPHOTOMETRIC 
UNCERTAINTY}
\label{sec:sperr}
In this section, we quantify the spectrophotometric errors in 
the SDSS-RM survey. The spectra were calibrated using $70$ 
standard stars. Therefore, the variability of standard stars 
used in the SDSS-RM survey should reflect the spectrophotometric 
errors. Note that one star was found to be intrinsically 
variable, and five other stars have an epoch with a ``dropped'' 
spectrum. We do not remove these stars when accounting for the 
variability of standard stars since these stars were not removed 
when calibrating the spectra. Due to their small number, these 
stars do not significantly affect the flux calibration.

\subsection{STRUCTURE FUNCTION OF 
STANDARD STARS}
\label{sec:sfstar}
As a first step, we present the observed distribution 
of $\Delta m$ for standard stars. To do so, we created 
$21$ wavelength bins starting from $3700\ \rm{\AA}$, 
each with a width of $300\ \rm{\AA}$. We then 
calculated the average flux in each bin at every 
epoch for every standard star. As a final step, we 
again calculated $\Delta m$ for any two observations 
separated by $\Delta t$. 

\begin{figure}
\epsscale{1.2}
\plotone{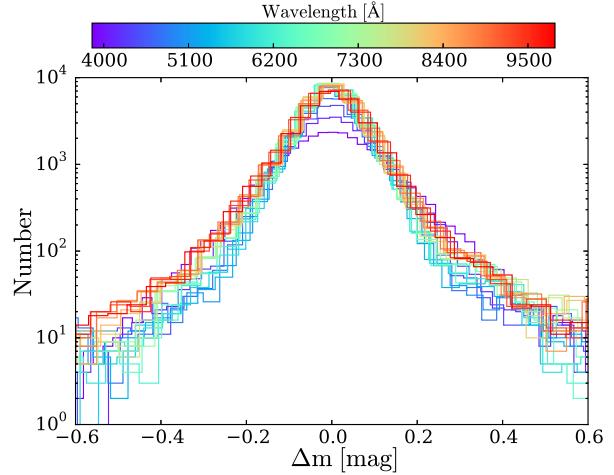}
\caption{The distributions of $\Delta m$ for standard 
stars. Again, $\Delta m=-2.5\log (f_2/f_1)$. There are 
exponential tails in the distribution of $\Delta m$ 
for standard stars. The variability is the smallest at 
$\sim 6000-7000\ \rm{\AA}$ (since the flux calibration 
is done using the $r$-band), and higher at both small 
and large wavelengths. }
\label{fig:var-star}
\end{figure}

In Figure~\ref{fig:var-star}, we plot the distribution 
of $\Delta m$ for standard stars. It is evident that 
the variability of standard stars depends strongly on 
wavelength. The variability at $\sim 7000\ \rm{\AA}$ 
is the smallest ($\sim 0.06\ \rm{mag}$). Also, the 
variability increases significantly to both the 
short-wavelength ($\sim 0.1\ \rm{mag}$) and the 
long-wavelength ends ($\sim 0.07\ \rm{mag}$). This 
is due to the fact that the flux calibration is done 
using $r$-band only. Note that the distributions are 
not Gaussian but are instead Laplacian which indicates 
that the spectrophotometric errors are not perfectly 
Gaussian (and the tails are not due to the variable 
stars). 

\begin{figure}
\epsscale{1.2}
\plotone{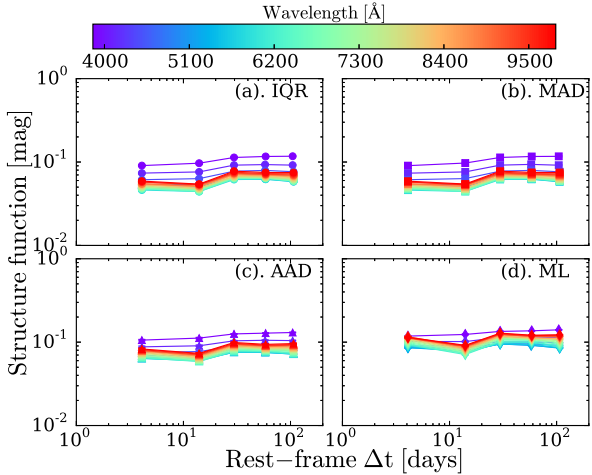}
\caption{The structure function of standard stars 
for each wavelength bin. The spectrophotometric 
uncertainty of a single epoch is the structure function 
divided by $\sqrt{2}$. The spectrophotometric errors 
depend on wavelength and have a minimum value 
at $6000-7000\ \rm{\AA}$ (of $\sim 0.06/\sqrt{2}\ \rm{mag}$). 
Furthermore, the structure function only slightly depends 
on $\Delta t$ (increasing by $0.02\ \rm{mag}$ at 
$\Delta t \gtrsim 20$ days). }
\label{fig:sfstar}
\end{figure}

Since the quasar spectra were calibrated using standard 
stars, we can then use the structure function of 
standard stars as an estimation of spectrophotometric 
errors. \textit{Note that the structure function quantifies 
the variability of a flux pair, and therefore the 
spectrophotometric uncertainty for a single epoch is 
the structure function divided by $\sqrt{2}$}. 
The variability of standard stars depends on wavelength, 
and so we calculated the structure function of 
standard stars in each wavelength bin separately. 
Figure~\ref{fig:sfstar} plots our results. It is evident 
that the structure function of the standard stars is 
largely constant with $\Delta t$ for each wavelength bin. 
There is a small increase (within $0.02\ \rm{mag}$) at 
$\Delta t>20\ \rm{days}$: this is associated with the 
time between different dark runs in the 
SDSS-RM observations. The structure function of standard 
stars depends on wavelength and is the smallest in the 
$\sim 6000-7000\ \rm{\AA}$ band. In the $\sim 6000-7000\ 
\rm{\AA}$ band, the scale of the variability is only $\sim 
0.06\ \rm{mag}$ (for the IQR estimator). Therefore, for a 
single epoch, the spectrophotometric error of our SDSS-RM 
data is expected to be not smaller than $\sim 0.06/\sqrt{2} 
\approx 0.04\ \rm{mag}$ \citep[also see Figure~20 of][]{shen15}. 

The spectrophotometric error of the SDSS-I/II data for 
a single epoch is assumed to be $0.04\ \rm{mag}$ \citep{dr6}.

\subsection{\rm{\O3}}
\label{sec:o3}
We also checked the variability of \O3$\lambda5007$\ 
(hereafter \O3). Physically, we expect that there is no 
intrinsic variability of \O3 \ on timescales of $\sim 100\ 
\rm{days}$. That is, the variability of \O3 should be 
equivalent to that of standard stars. To verify this, we 
calculated the structure function of \O3 \ while subtracting 
only measurement errors (i.e., $\sigma_{\rm{e}}$ 
in Section~\ref{sec:def} only accounts for measurement 
errors) and compared the structure functions with the expected 
spectrophotometric errors of the flux pairs. The expected 
spectrophotometric errors are the average of the structure 
function of standard stars at the same wavelength as the 
\O3 \ (i.e., $5007(1+z)\ \rm{\AA}$, where $z$ is the 
redshift of the quasar). We divided our sources into 
subsamples of point sources and extended sources 
in order to investigate the additional variability 
due to variable seeing. To investigate the effects of 
measurement errors on the estimation of the structure 
function, we binned quasars by $\widetilde{\rm{S/N}}$. 
For each sub-sample, we created three bins: $15 \leq 
\widetilde{\rm{S/N}}<25$, $25\leq \widetilde{\rm{S/N}}<35$, 
$\widetilde{\rm{S/N}}\geq 35$. We required a minimum 
of $\widetilde{\rm{S/N}} > 15$ so that the measurement 
errors do not dominate over the spectrophotometric 
errors. We then calculated the structure function of 
each bin. 

\begin{figure}
\epsscale{1.2}
\plotone{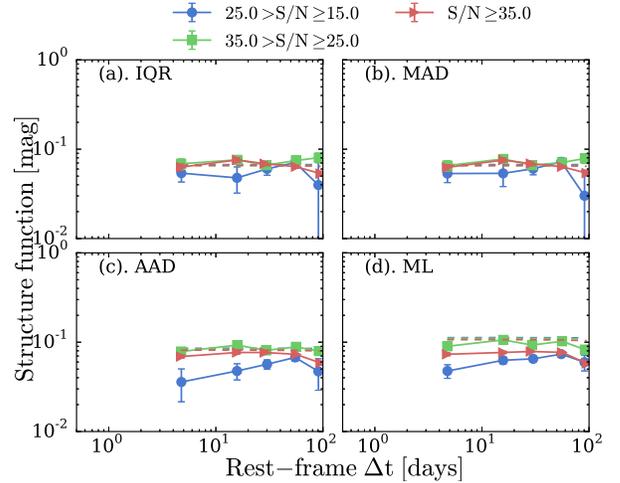}
\caption{The structure functions of \O3 for point sources. 
Note that the structure functions presented here 
were only corrected for measurement errors. The dashed 
lines represent the expected spectrophotometric errors 
which are obtained according to the redshift distribution 
of quasars in each bin. It is evident that, for point sources, 
\O3 does not show intrinsic variability. This result indicates 
that our estimation of the spectrophotometric errors is robust. 
}
\label{fig:o3-point}
\end{figure}

Figure~\ref{fig:o3-point} shows the structure function for 
the sub-sample of point sources. The variability of [OIII] 
for point sources is nearly identical to the expected 
spectrophotometric errors. For each bin, we calculated the 
difference between the structure function and the 
spectrophotometric errors in quadrature. The 
difference is no more than $0.03$ mag, and there is no 
systematic offset. In addition, the variability is not a 
monotonic function of $\widetilde{\rm{S/N}}$ which suggests 
our $\widetilde{\rm{S/N}}>15$ cut reliably prevents a bias 
from large measurement errors dominating the observed 
variability. \textit{Therefore, we conclude that our 
estimation of the spectrophotometric errors is robust}. 

\begin{figure}
\epsscale{1.2}
\plotone{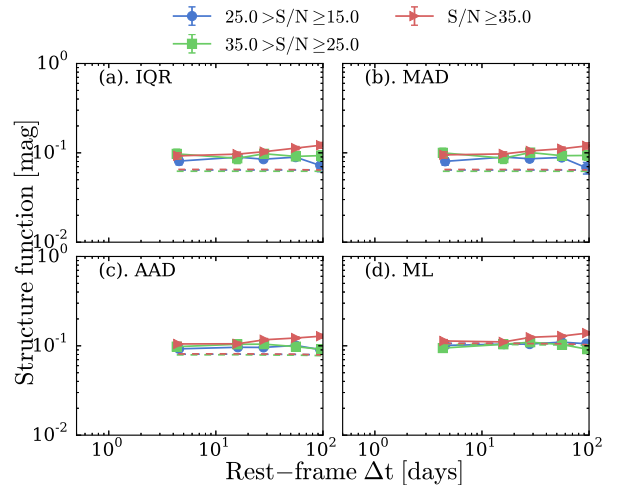}
\caption{The structure functions of \O3 for extended sources 
only. Note that the structure functions that are presented here 
were only corrected for measurement errors. The dashed lines 
represent the expected spectrophotometric errors which are 
obtained according to the redshift distribution of quasars in 
each bin. Extended sources show variability beyond the 
spectrophotometric errors which is likely due to seeing 
effects (i.e., the variable seeing in different epochs).}
\label{fig:o3-extend}
\end{figure}

Figure~\ref{fig:o3-extend} plots the structure function for 
the sub-sample of extended sources. It is evident that, for 
each bin, \O3 \ shows excess variability compared to the 
spectrophotometric errors. This is due to additional variability 
($\sim 0.07-0.1\ \rm{mag}$ in the IQR) from variable seeing 
in addition to the spectrophotometric errors. Additional 
variability from seeing changes has consequences for the 
structure-function estimates of extended sources (e.g., if 
there is a substantial host-galaxy component to the measured 
$5100\ \rm{\AA}$ luminosity, see Section~\ref{sec:sf5k}). 

\clearpage


\clearpage
\begin{thebibliography}{}
\bibitem[Adelman-McCarthy et al.(2008)]{dr6} 
Adelman-McCarthy, J.~K., Ag{\"u}eros, M.~A., Allam, S.~S., et al.\ 2008, 
\apjs, 175, 297  

\bibitem[Ai et al.(2010)]{ai10} Ai, Y.~L., Yuan, W., Zhou, 
H.~Y., et al.\ 2010, \apjl, 716, L31 

\bibitem[Almaini et al.(2000)]{alm00} Almaini, O., Lawrence, 
A., Shanks, T., et al.\ 2000, \mnras, 315, 325   

\bibitem[Bauer et al.(2009)]{bau09}
Bauer, A., Baltay, C., Coppi, P., Ellman, N., Jerke, J., Rabinowitz,
D. \& Scalzo, R. 2009, ApJ, 696, 1241

\bibitem[Blandford \& McKee(1982)]{bla82}
Blandford, R. D. \& McKee, C. F. 1982, ApJ, 255, 419

\bibitem[Bolton et al.(2012)]{bol12} Bolton, A.~S., Schlegel, 
D.~J., Aubourg, {\'E}., et al.\ 2012, \aj, 144, 144 

\bibitem[Cackett et al.(2015)]{cac15} Cackett, E.~M., 
Gultekin, K., Bentz, M.~C., et al.\ 2015, arXiv:1503.02029 

\bibitem[Clavel et al.(1991)]{cla91} Clavel, J., Reichert, 
G.~A., Alloin, D., et al.\ 1991, \apj, 366, 64

\bibitem[Czerny(2006)]{czerny06} Czerny, B.\ 2006, Astronomical 
Society of the Pacific Conference Series, 360, 265 

\bibitem[Czerny et al.(1999)]{czerny99}
Czerny, B., Schwarzenberg-Czerny, A. \& and Loska, Z. 1999, MNRAS,
303, 148

\bibitem[Dawson et al.(2013)]{daw13} Dawson, K.~S., Schlegel, 
D.~J., Ahn, C.~P., et al.\ 2013, \aj, 145, 10 

\bibitem[De Rosa et al.(2015)]{de15} De Rosa, G., Peterson, 
B.~M., Ely, J., et al.\ 2015, \apj, 806, 128 

\bibitem[de Vries et al.(2005)]{dv05}
de Vries, W. H., Becker, R. H., White, R. L. \& Loomis, C. 2005, AJ,
129, 615

\bibitem[Eisenstein et al.(2011)]{eis11} Eisenstein, D.~J., 
Weinberg, D.~H., Agol, E., et al.\ 2011, \aj, 142, 72 

\bibitem[Emmanoulopoulos et al.(2010)]{emm10} 
Emmanoulopoulos, D., McHardy, I.~M., \& Uttley, P.\ 2010, \mnras, 404, 931 

\bibitem[Gaskell(2009)]{gas09} Gaskell, C.~M.\ 2009, New Astron. Rev., 53, 140 

\bibitem[Giveon et al.(1999)]{giv99}
Giveon, U., Maoz, D., Kaspi, S., Netzer, H. \& Smith, P.  S. 1999,
MNRAS, 306, 637

\bibitem[Grier et al.(2013)]{gri13} Grier, C.~J., Peterson, 
B.~M., Horne, K., et al.\ 2013, \apj, 764, 47

\bibitem[Gunn et al.(2006)]{gun06} Gunn, J.~E., Siegmund, 
W.~A., Mannery, E.~J., et al.\ 2006, \aj, 131, 2332

\bibitem[Halpern et al.(2003)]{hal03} Halpern, J.~P., 
Leighly, K.~M., \& Marshall, H.~L.\ 2003, \apj, 585, 665

\bibitem[Hawkins(2002)]{haw02}
Hawkins, M. R. S. 2002, MNRAS, 329, 76

\bibitem[Hirose et al.(2009)]{hir09} Hirose, S., Blaes, O., 
\& Krolik, J.~H.\ 2009, \apj, 704, 781 

\bibitem[Hook et al.(1994)]{hook94}
Hook, I. M., McMahon, R. G., Boyle, B. J. \& Irwin, M. J. 1994,
MNRAS, 268, 305

\bibitem[Hryniewicz et al.(2014)]{hry14} Hryniewicz, K., 
Czerny, B., Pych, W., et al.\ 2014, \aap, 562, AA34

\bibitem[Ivezic et al.(2004)]{ive04} Ivezic, {\v Z}., Lupton, 
R.~H., Juric, M., et al.\ 2004, in IAU Symp. 222, The Interplay 
Among Black Holes, Stars and ISM in Galactic Nuclei, ed. Th. 
Storchi Bergmann, L.C. Ho, \& H. R. Schmitt (Cambridge: Cambridge 
Univ. Press), 525 

\bibitem[Jiang et al.(2013)]{jiang13} Jiang, Y.-F., Stone, 
J.~M., \& Davis, S.~W.\ 2013, \apj, 778, 65 

\bibitem[Kasliwal et al.(2015)]{kas15} Kasliwal, V.~P., 
Vogeley, M.~S., \& Richards, G.~T.\ 2015, arXiv:1505.00360 

\bibitem[Kaspi et al.(2000)]{kas00} Kaspi, S., Smith, P.~S., 
Netzer, H., et al.\ 2000, \apj, 533, 631 

\bibitem[Kato et al.(1998)]{kat98} Kato, S., Fukue, J., 
\& Mineshige, S.\ 1998, Black-hole Accretion Disks (Kyoto: 
Kyoto University Press) 

\bibitem[Kelly et al.(2009)]{kelly09}
Kelly, B. C., Bechtold, J. \& Siemiginowska, A. 2009, ApJ, 698, 895

\bibitem[Kelly et al.(2011)]{kelly11} Kelly, B.~C., Sobolewska, 
M., \& Siemiginowska, A.\ 2011, \apj, 730, 52

\bibitem[Kokubo et al.(2014)]{kok14}
Kokubo, M., Morokuma, T., Minezaki, T., Doi, M., Kawaguchi, T.,
Sameshima, H. \& Koshida, S. 2014, ApJ, 783, 46

\bibitem[Korista et al.(1997)]{kor97} Korista, K., Baldwin, 
J., Ferland, G., \& Verner, D.\ 1997, \apjs, 108, 401 

\bibitem[Korista 
\& Goad(2000)]{kor00} Korista, K.~T., \& Goad, M.~R.\ 2000, \apj, 536, 284

\bibitem[Korista 
\& Goad(2004)]{kor04} Korista, K.~T., \& Goad, M.~R.\ 2004, \apj, 606, 749 

\bibitem[Krawczyk et al.(2013)]{kra13} Krawczyk, C.~M., 
Richards, G.~T., Mehta, S.~S., et al.\ 2013, \apjs, 206, 4 

\bibitem[Laor(1998)]{lao98} Laor, A.\ 1998, \apjl, 505, L83

\bibitem[Li \& Cao(2008)]{li08} Li, S.-L., \& Cao, X.\ 2008, \mnras, 387, L41 

\bibitem[Lin et al.(2012)]{lin12} Lin, D.-B., Gu, W.-M., Liu, 
T., Sun, M.-Y., \& Lu, J.-F.\ 2012, \apj, 761, 29 

\bibitem[Lusso et al.(2012)]{lus12} Lusso, E., Comastri, A., 
Simmons, B.~D., et al.\ 2012, \mnras, 425, 623

\bibitem[Lyubarskii(1997)]{lyu97} Lyubarskii, Y.~E.\ 1997, 
\mnras, 292, 679 

\bibitem[MacAlpine(1972)]{mac72} MacAlpine, G.~M.\ 1972, 
\apj, 175, 11

\bibitem[MacLeod et al.(2008)]{mac08}
MacLeod, C. L., Ivezi\'{c}, \v{Z}., de Vries, W., Sesar, B. \&
Becker, A. 2008, AIPC, 1082, 282

\bibitem[MacLeod et al.(2010)]{mac10}
MacLeod, C. L., Ivezi\'{c}, \v{Z}., Kochanek, C. S. et al. 2010, ApJ,
721, 1014 (M10)

\bibitem[MacLeod et al.(2012)]{mac12}
MacLeod, C. L., Ivezi\'{c}, \v{Z}., Sesar, B. et al. 2012, ApJ, 753,
106

\bibitem[Margala et al.(2015)]{mar15} Margala, D., Kirkby, D., Dawson, K., 
et al.\ 2015, in preparation  

\bibitem[Marshall et al.(1997)]{mar97} Marshall, H.~L., 
Carone, T.~E., Peterson, B.~M., et al.\ 1997, \apj, 479, 222 

\bibitem[Morgan et al.(2010)]{mor10} Morgan, C.~W., Kochanek, 
C.~S., Morgan, N.~D., \& Falco, E.~E.\ 2010, \apj, 712, 1129 

\bibitem[Mushotzky et al.(2011)]{mus11}
Mushotzky, R. F., Edelson, R., Baumgartner, W. \& Gandhi, P. 2011,
ApJ, 743, 12

\bibitem[Netzer(1980)]{net80} Netzer, H.\ 1980, \apj, 236, 
406 

\bibitem[Peterson(1993)]{pet93} Peterson, B.~M.\ 1993, \pasp, 
105, 247

\bibitem[Peterson(2014)]{pet14} Peterson, B.~M.\ 2014, \ssr, 
183, 253

\bibitem[Peterson et al.(1998)]{pet98} Peterson, B.~M., 
Wanders, I., Bertram, R., et al.\ 1998, \apj, 501, 82 

\bibitem[Peterson \& Bentz(2006)]{pet06}
Peterson, B. M. \& Bentz, M. C. 2006, NewAR, 50, 796

\bibitem[Reichert et al.(1994)]{rei94} Reichert, G.~A., 
Rodriguez-Pascual, P.~M., Alloin, D., et al.\ 1994, \apj, 425, 582 

\bibitem[Richards et al.(2006)]{ric06} Richards, G.~T., Lacy, 
M., Storrie-Lombardi, L.~J., et al.\ 2006, \apjs, 166, 470 

\bibitem[Sesar et al.(2007)]{ses07} Sesar, B., Ivezi{\'c}, 
{\v Z}., Lupton, R.~H., et al.\ 2007, \aj, 134, 2236

\bibitem[Shakura \& Sunyaev(1973)]{sha73} Shakura, N.~I., \& 
Sunyaev, R.~A.\ 1973, \aap, 24, 337 

\bibitem[Shen(2013)]{shen13} Shen, Y.\ 2013, BASI, 41, 61

\bibitem[Shen et al.(2015a)]{shen15}
Shen, Y., Brandt, W. N., Dawson, K. S. et al. 2015a, ApJS, 216, 4

\bibitem[Shen et al.(2015b)]{shen15b} Shen, Y., Greene, J.~E., 
Ho, L.~C., et al.\ 2015b, \apj, 805, 96

\bibitem[Shen et al.(2015c)]{shen15c} Shen, Y., Horne, K., et al.\ 2015c, in preparation 

\bibitem[Shen \& Liu(2012)]{shen12} Shen, Y., \& Liu, X.\ 2012, \apj, 753, 125

\bibitem[Shen et al.(2008)]{shen08} Shen, Y., Greene, J.~E., 
Strauss, M.~A., Richards, G.~T., \& Schneider, D.~P.\ 2008, \apj, 680, 169

\bibitem[Shen et al.(2011)]{shen11} Shen, Y., Richards, G.~T., 
Strauss, M.~A., et al.\ 2011, \apjs, 194, 45 

\bibitem[Smee et al.(2013)]{sme13} Smee, S.~A., Gunn, J.~E., 
Uomoto, A., et al.\ 2013, \aj, 146, 32 

\bibitem[Trevese et al.(2007)]{tre07} Trevese, D., Paris, D., 
Stirpe, G.~M., Vagnetti, F., \& Zitelli, V.\ 2007, \aap, 470, 491 

\bibitem[Ulrich et al.(1997)]{ulrich97}
Ulrich, M.-H., Maraschi, L. \& Urry, C. M. 1997, ARA\&A, 35, 445

\bibitem[Uomoto et al.(1976)]{uom76}
Uomoto, A. K., Wills, B. J. \& Wills, D. 1976, AJ, 81, 905

\bibitem[Uttley et al.(2000)]{utt00} Uttley, P., McHardy, 
I.~M., Papadakis, I.~E., Cagnoni, I., 
\& Fruscione, A.\ 2000, \mnras, 312, 880 

\bibitem[Vanden Berk et al.(2004)]{vb04}
Vanden Berk, D. E., Wilhite, B. C., Kron, R. G. et al. 2004, ApJ,
601, 692

\bibitem[Vestergaard 
\& Osmer(2009)]{ves09} Vestergaard, M., \& Osmer, P.~S.\ 2009, \apj, 699, 800

\bibitem[Vestergaard \& Peterson(2006)]{ves06} Vestergaard, M., 
\& Peterson, B.~M.\ 2006, \apj, 641, 689

\bibitem[Wang et al.(2009)]{wan09} Wang, J.-G., Dong, X.-B., 
Wang, T.-G., et al.\ 2009, \apj, 707, 1334

\bibitem[Wilhite et al.(2008)]{wil08} Wilhite, B.~C., 
Brunner, R.~J., Grier, C.~J., Schneider, D.~P., 
\& vanden Berk, D.~E.\ 2008, \mnras, 383, 1232 


\bibitem[Woo(2008)]{woo08}
Woo, J.-H. 2008, AJ, 135, 1849

\bibitem[York et al.(2000)]{sdss}
York, D.~G. et al.~2000, AJ, 120, 1579

\bibitem[Zu et al.(2013)]{zu13} Zu, Y., Kochanek, C.~S., 
Koz{\l}owski, S., \& Udalski, A.\ 2013, \apj, 765, 106 

\bibitem[Zuo et al.(2012)]{zuo12} Zuo, W., Wu, X.-B., Liu, 
Y.-Q., \& Jiao, C.-L.\ 2012, \apj, 758, 104 

\end{thebibliography}
\end{document}